\documentclass[twocolumn]{aastex631}

\usepackage{amsmath}
\usepackage{subfigure}

\shorttitle{ALPINE radio stacking and AGN fraction}
\shortauthors{Shen et al.}
\graphicspath{{./}{figures/}}

\begin{document}
\title{The ALPINE-ALMA [CII] survey: \\ The infrared-radio correlation and AGN fraction of star-forming galaxies at z $\sim$ 4.4 -- 5.9 }


\correspondingauthor{Lu Shen}
\email{lushen@ustc.edu.cn}

\author[0000-0001-9495-7759]{Lu Shen}
\affil{CAS Key Laboratory for Research in Galaxies and Cosmology, Department of Astronomy, University of Science and Technology of China, Hefei 230026, China}
\affil{School of Astronomy and Space Sciences, University of Science and Technology of China, Hefei, 230026, China}

\author[0000-0002-1428-7036]{Brian C. Lemaux}
\affil{Gemini Observatory, NSF's NOIRLab, 670 N. A'ohoku Place, Hilo, Hawai'i, 96720, USA}
\affil{Department of Physics and Astronomy, University of California, Davis, One Shields Ave., Davis, CA 95616, USA}

\author{Lori M. Lubin} 
\affil{Department of Physics and Astronomy, University of California, Davis, One Shields Ave., Davis, CA 95616, USA}

\author{Guilin Liu}
\affil{CAS Key Laboratory for Research in Galaxies and Cosmology, Department of Astronomy, University of Science and Technology of China, Hefei 230026, China}
\affil{School of Astronomy and Space Sciences, University of Science and Technology of China, Hefei, 230026, China}

\author{Matthieu B\'{e}thermin}
\affil{Aix-Marseille Univ, CNRS, CNES, Laboratoire d'Astrophysique de Marseille, Marseille, France}

\author{M\'{e}d\'{e}ric Boquien}
\affil{Centro de Astronomía (CITEVA), Universidad de Antofagasta, Avenida Angamos 601, Antofagasta, Chile}

\author{Olga Cucciati}
\affil{INAF - Osservatorio di Astrofisica e Scienza dello Spazio di Bologna, via Gobetti 93/3 - 40129 Bologna - Italy}

\author{Olivier Le F\`{e}vre}
\affil{Aix-Marseille Univ, CNRS, CNES, Laboratoire d'Astrophysique de Marseille, Marseille, France}

\author{Margherita Talia}
\affil{University of Bologna - Department of Physics and Astronomy ``Augusto Righi'' (DIFA), Via Gobetti 93/2, I-40129, Bologna, Italy}

\author{Daniela Vergani}
\affil{INAF - Osservatorio di Astrofisica e Scienza dello Spazio di Bologna, via Gobetti 93/3 - 40129 Bologna - Italy}

\author{Gianni Zamorani}
\affil{INAF - Osservatorio di Astrofisica e Scienza dello Spazio, Via Gobetti 93/3, I-40129, Bologna, Italy}

\author{Andreas L. Faisst}
\affil{IPAC, M/C 314-6, California Institute of Technology, 1200 East California Boulevard, Pasadena, CA 91125, USA}

\author{Michele Ginolfi}
\affil{Observatoire de Gen\`{e}ve, Universit\`{e} de Gen\`{e}ve, 51 Ch. des Maillettes, 1290 Versoix, Switzerland}

\author{Carlotta Gruppioni}
\affil{INAF - Osservatorio di Astrofisica e Scienza dello Spazio di Bologna, via Gobetti 93/3, I-40129, Bologna, Italy}

\author{Gareth C. Jones}
\affil{Cavendish Laboratory, University of Cambridge, 19 J. J. Thomson Ave., Cambridge CB3 0HE, UK}
\affil{Kavli Institute for Cosmology, University of Cambridge, Madingley Road, Cambridge CB3 0HA, UK}
\affil{Cavendish Laboratory/ Kavli Institute for Cosmology, University of Cambridge, 19 J. J. Thomson Ave., Cambridge CB3 0HE, UK}
\affil{Department of Physics, University of Oxford, Denys Wilkinson Building, Keble Road, Oxford OX1 3RH, UK}

\author{Sandro Bardelli}
\affil{INAF - Osservatorio di Astrofisica e Scienza dello Spazio di Bologna, via Gobetti 93/3 - 40129 Bologna - Italy}

\author{Nimish Hathi}
\affil{Space Telescope Science Institute, 3700 San Martin Drive, Baltimore, MD 21218, USA}

\author{Anton M. Koekemoer}
\affil{Space Telescope Science Institute, 3700 San Martin Drive, Baltimore, MD 21218, USA}

\author{Michael Romano}
\affil{National Centre for Nuclear Research, ul.Pasteura 7, 02-093 Warsaw, Poland}
\affil{Dipartimento di Fisica e Astronomia, Universit\`{a} di Padova, Vicolo dell'Osservatorio 3, I-35122, Padova, Italy}
\affil{INAF - Osservatorio Astronomico di Padova, Vicolo dell'Osservatorio 5, I-35122, Padova, Italy}

\author{Daniel Schaerer}
\affil{11 Observatoire de Gen`eve, Universit\`{e} de Gen\`{e}ve, 51 Ch. des Maillettes, 1290 Versoix, Switzerland}

\author{Elena Zucca}
\affil{INAF - Osservatorio di Astrofisica e Scienza dello Spazio di Bologna, via Gobetti 93/3 - 40129 Bologna - Italy}

\author{Wenjuan Fang}
\affil{CAS Key Laboratory for Research in Galaxies and Cosmology, Department of Astronomy, University of Science and Technology of China, Hefei 230026, China}
\affil{School of Astronomy and Space Sciences, University of Science and Technology of China, Hefei, 230026, China}

\author{Ben Forrest} 
\affil{Department of Physics and Astronomy, University of California, Davis, One Shields Ave., Davis, CA 95616, USA}

\author{Roy Gal}
\affil{University of Hawai'i, Institute for Astronomy, 2680 Woodlawn Drive, Honolulu, HI 96822, USA}

\author{Denise Hung} 
\affil{University of Hawai'i, Institute for Astronomy, 2680 Woodlawn Drive, Honolulu, HI 96822, USA}

\author{Ekta A. Shah} 
\affil{Department of Physics and Astronomy, University of California, Davis, One Shields Ave., Davis, CA 95616, USA}

\author{Priti Staab} 
\affil{Department of Physics and Astronomy, University of California, Davis, One Shields Ave., Davis, CA 95616, USA}

\author{Brittany Vanderhoof}
\affil{School of Physics and Astronomy, Rochester Institute of Technology, Rochester, NY 14623, USA}

\author{Eduardo Ibar}
\affil{Instituto de F\'isica y Astronom\'ia, Universidad de Valpara\'iso, Avda. Gran Breta\~na 1111, Valpara\'iso, Chile}

\begin{abstract}

We present the radio properties of 66 spectroscopically-confirmed normal star-forming galaxies (SFGs) at $4.4<z<5.9$ in the COSMOS field that were [C \textsc{ii}] detected in the Atacama Large Millimeter Array (ALMA) Large Program to INvestigate [C \textsc{ii}] at Early times (ALPINE). 
We separate these galaxies (``CII-detected-all'') into lower redshift (``CII-detected-lz'', $\langle z\rangle=4.5$) and higher redshift (``CII-detected-hz'', $\langle z\rangle=5.6$) sub-samples and stack multi-wavelength imaging for each sub-sample from X-ray to radio bands. 
A radio signal is detected in the stacked 3 GHz image of CII-detected-all and -lz samples at $\gtrsim3\sigma$. 
We find that the infrared-radio correlation of our sample, quantified by $q_{\mathrm{TIR}}$, is lower than the local relation for normal SFGs at $\sim$3$\sigma$ significance level, and is instead broadly consistent with that of bright sub-mm galaxies at $2<z<5$. 
Neither of these samples show evidence of dominant AGN activity in their stacked Spectral Energy Distributions (SEDs), rest-frame UV spectra, or X-ray images. 
Although we cannot rule out the possible effect of the assumed spectral index and the applied infrared SED templates as at least partially causing these differences, the lower obscured fraction of star formation than at lower redshift can alleviate the tension between our stacked $q_{\mathrm{TIR}}$s and that of local normal SFGs. It is possible that the dust buildup, which primarily governs the IR emission in addition to older stellar populations, has not had enough time to occur fully in these galaxies, whereas the radio emission can respond on a more rapid timescale. Therefore, we might expect a lower $q_{\mathrm{TIR}}$ to be a general property of high-redshift SFGs.

\end{abstract}

\keywords{Radio galaxies (1343); high-redshift – submillimeter: galaxies; Galaxy evolution (594); }


\section{Introduction} \label{sec:intro}

A tight correlation between the total infrared (IR) and radio luminosity of star-forming galaxies (SFGs) has been known for decades, initially found to be hold over three order of magnitude in both IR and radio luminosity (e.g. \citealp{deJong1985, Helou1985, Condon1992, Yun2001, Bell2003}). 
Early observations have concluded that IR-radio correlation (IRRC) does not appear to evolve over at least the past 10–12 Gyr of cosmic history (e.g. \citealp{Ibar2008, Sajina2008, Murphy2009, Sargent2010, Garrett2002, Appleton2004, Garn2009, Jarvis2010, Mao2011, Bourne2011, Smith2014}). The IRRC is commonly quantified by the logarithmic ratio of the total infrared (8 - 1000 $\mu$m) and 1.4 GHz luminosities ($q_\mathrm{{TIR}}$). 
Based on this tight IRRC, radio luminosity is well calibrated and used as a star formation rate (SFR) indicator (e.g. \citealp{Condon1992, Bell2003}). Meanwhile, the excess of radio luminosity as compared to this IRRC has been used to identify active galactic nuclei (AGN; e.g. \citealp{Donley2005, DelMoro2013, Lemaux2014b, Shen2017}) 

The star-formation activity is known to be responsible for the existence of the IRRC, although the detailed physical mechanism and process remain unclear (e.g., \citealp{Voelk1989, Bell2003, Lacki2010, Helou1993}). The IR originates from young, massive stars that emit ultraviolet (UV) photons, which are absorbed by dust and re-emitted in the IR region. The process is more complicated for the radio continuum emission which originates from two components: (1) thermal emission from free-free interactions of ionized particles due to H II regions surrounding recently formed high-mass stars, and (2) non-thermal synchrotron emission due to supernova of these stars after $\sim$10 Myr (e.g., \citealp{Condon1992}). Whereas, the radio continuum emission is expected to be modified at high redshift, due to the increasing typical interstellar medium (ISM) density and stronger magnetic fields, which in turn enhance the radio emission \citep{Schleicher2013}. In addition, the smaller dust attenuation factor observed in high-redshift galaxies may affect the IRRC in terms of decreasing the expected IR emission (e.g., \citealp{Carilli2008}). 
Nevertheless, the contribution of AGN to IR emission (e.g. \citealp{Mor2012, Symeonidis2016, Mullaney2012, Lyu2017, LyuRieke2017}) and to radio luminosity (e.g., \citealp{Condon1992, DelMoro2013}) is known to affect the IRRC, shown by the lower $q_{\mathrm{TIR}}$ value. 

More recent studies have found a moderate, but statistically significant evolution of the IRRC \citep{Magnelli2015, Ivison2010, Casey2012, Delhaize2017, CalistroRivera2017}. 
In particular, \citet{Delhaize2017} have well quantified the evolution of $q_{\mathrm{TIR}}$ out to $z\sim 3$ for SFGs using the deep 3 GHz data of  \citet{Smolcic2017} and Herschel data \citep{Lutz2011, Oliver2012} in the COSMOS field. 
These results agree with the expectation of the evolution of IRRC with an enhanced radio emission and/or a decreased IR emission at $z\sim3$. 
However, studies have found that the $q_{\mathrm{TIR}}$ depends on radio luminosity \citep{Molnar2021}, stellar mass \citep{Gurkan2018, Smith2021, Delvecchio2021}, and observed radio frequency due to the different values of radio slope \citep{An2021}, with the dependence of these being stronger than that on redshift.

While attempts to quantify the IRRC for high-z SFGs (z $>$ 3) have been made, past studies have only investigated SMGs \citep{Murphy2009, Michalowski2010, Thomson2014, Miettinen2017, Algera2020, Smolcic2015}. These studies found that SMGs at high redshift (z $>$ 2) generally lie significantly below the canonical $q_{\mathrm{TIR}}$ ratio of local normal SFGs \citep{Murphy2009, Miettinen2017, Algera2020, Smolcic2015}. 
However, SMGs provide only a biased view of the intensely star-forming population with SFR $>$ 1000 M$_\odot$ yr$^{-1}$. 
Existing observations of radio and IR for normal SFGs, lying on the SFR- M$_*$ relation for SFGs at these redshifts (e.g., \citealp{Speagle2014, Tasca2015, Tomczak2016, Pearson2018, Khusanova2020}), have not been conducted individually or statistically. 
The ALMA Large Program to Investigate C$^+$ at Early Times (ALPINE) survey provides measurements on the rest-frame far-infrared (FIR) continuum emission for a representative sample of 118 normal SFGs at $4.4 < z < 5.9$. 
Using this data set combined with the available deep radio observations for most of these galaxies, we investigate here the IRRC of normal high-z SFGs via a stacking analysis. 
In this paper, we present the ancillary data available for our sample and sample selection in Section \ref{sec:data}. 
In Section \ref{sec:methods}, we describe the stacking method and the spectral energy distribution (SED) fitting, which estimates the AGN contribution to the total IR luminosity (f$_{\rm{AGN}}$). 
Our results on the radio detection, the $q_{\mathrm{TIR}}$ and the fraction of AGN component are presented in Section \ref{sec:result}. 
We discuss the possible explanation for the lower $q_{\mathrm{TIR}}$ in Section \ref{sec:discussion}. 
We conclude with a summary in Section \ref{sec:summary}. 
In this paper, we assume Chabrier (2003) initial mass function (IMF) and a $\Lambda$CDM cosmology with $\Omega_\Lambda$ = 0.7, $\Omega_\mathrm{m}$ = 0.3, and $H_0$ = 70 km s$^{-1}$ Mpc$^{-1}$.

\section{Data and Sample Selection} \label{sec:data}

\subsection{ALPINE}

ALPINE is an ALMA large program (Project ID: 2017.1.00428.L; PI: O. Le F\`{e}vre; see also e.g., \citealp{LeFevre2020, Faisst2020, Bethermin2020}), aimed at measuring [C \textsc{ii}] 158 $\mu$m and rest-frame FIR continuum emission from a representative sample of 118 main-sequence galaxies at $4.4 < z < 5.9$. 
All galaxies are selected to be spectroscopically confirmed with rest-frame UV spectroscopy from the VIMOS Ultra Deep Survey (VUDS, \citealp{LeFevre2015, Tasca2017}) and COSMOS 10K survey \citep{Capak2004, Mallery2012, Hasinger2018} in the redshift range of $4.3 < z_{\mathrm{spec}} < 4.6$ and $5.1 < z_{\mathrm{spec}} < 5.9$. 
The gap in redshift range is due to the presence of an atmospheric absorption feature.  
Galaxies were selected to have SFR$_\mathrm{SED}$ $\ge$ 10 M$_*$ yr$^{-1}$ from the Cosmic Evolution Survey (COSMOS, 105 galaxies, \citealp{Scoville2007a}) and the Extended Chandra Deep Field South (ECDFS, 13 galaxies, \citealp{Giacconi2002}). 
Broad-line AGN, identified by their rest-frame UV spectra, were not selected as ALPINE targets. One galaxy in the ALPINE sample was found to be detected at X-ray wavelengths indicating AGN activity, though it is not a Type-1 AGN (see section \ref{sec:xray} for more details). 
The construction and the physical properties of the sample are described in \citet{LeFevre2020, Faisst2020}. 
The ALMA observations and the data reduction are fully described in \citet{Bethermin2020}.

\subsection{Radio Data}\label{sec:radio_data}

The ALPINE galaxies in the COSMOS field were covered by the Very Large Array at 3 GHz from VLA-COSMOS 3 GHz Large Project \citep{Smolcic2017} and the 1.4 GHz in several campaigns \citep{Schinnerer2007, Schinnerer2010, Bertoldi2007}. 
The 3 GHz image was mapped with VLA S-band in the A and C configuration. The final 3 GHz image has a sensitivity of 2.3 $\mu$Jy beam$^{-1}$ and the synthesized beam is 0\arcsec.75. A total of 10,830 sources were detected above 5$\sigma$ using the \textsc{Blobcat} software.
The 1.4 GHz image was observed at VLA L-band in the A and C configuration where the synthesized beam is 1\arcsec.5 $\times$ 1\arcsec.4. The final 1.4 GHz image has a sensitivity of $\sim8\ \mu$Jy beam$^{-1}$ in the central 30$\arcmin \times$ 30$\arcmin$ and $\sim12\ \mu$Jy beam$^{-1}$ over the full area, respectively \citep{Karim2011}. 
Using the \textsc{aips} task \textsc{sad}, a total of 2865 sources were identified at $\ge5\sigma$ significance in the final VLA–COSMOS mosaic \citep{Schinnerer2010}.
The ALPINE galaxies in the ECDFS field are only observed by VLA at 1.4 GHz. The final image has the best sensitivity of 6 $\mu$Jy/beam. Using the \textsc{aips} task \textsc{sad} and \textsc{jmfit}, a total of 883 sources are identified at $\ge4\sigma$ significance \citep{miller2013}.

We cross match 3 GHz and 1.4 GHz sources to ALPINE galaxies using the 2\arcsec\ search radius, separately. 
Only one ALPINE galaxy (DEIMOS\_COSMOS\_842313) is matched to a 3 GHz source (COSMOSVLA3 J100054.49+023436.2) in the 2\arcsec\ search. 
It is known that DEIMOS\_COSMOS\_842313 has an extremely bright neighbor identified as SC\_2\_DEIMOS\_COSMOS\_842313 in \citet{Bethermin2020}, also known as J1000+0234 \citep{Schinnerer2008} or AzTEC/C17 \citep{Brisbin2017}. 
This 3 GHz source is closer to SC\_2\_DEIMOS\_COSMOS\_842313 with a separation of 0.12\arcsec, as compared to 1.17\arcsec\ to DEIMOS\_COSMOS\_842313, suggesting that the former is the match to the radio source. 
Thus, none of ALPINE galaxies are individually detected in the 3 GHz or 1.4 GHz image. 
In the absence of individual detections, we apply a stacking method, which is described in Section \ref{sec:stacking}, to investigate the statistical radio properties of ALPINE galaxies. 

\subsection{FIR Data} \label{sec:FIR}

The ALMA observations were conducted between 7 May, 2018 (Cycle 5) and 10 January, 2019 (Cycle 6) using antenna configurations C43-1 and C43-2. The integration times ranged from 15 to 45 minutes, with an average of 22 minutes. 
The [C \textsc{II}] intensity maps and FIR continuum maps at rest-frame 158 $\mu$m were produced using the line and line free channels, respectively (see details in \citealp{Bethermin2020}). 
The resulting median sensitivity of the continuum maps is 41 $\mu$Jy/beam in the range of 16.8 - 72.1 $\mu$Jy/beam. The average synthesized beam is 1\arcsec.13 $\times$ 0\arcsec.85.

\subsection{Optical/IR Imaging} \label{sec:optical}

The existing imaging data for ALPINE galaxies in the COSMOS and ECDFS field are summarized in \citet{Faisst2020}. 
Briefly, these include B, V, $g^+$, $r^+$, $i^+$, $z^{++}$ as well as 12 intermediate-band and 2 narrow-band filters from the Suprime-Cam on Subaru, the g, r, i, z, and y-band from the Hyper Suprime-Cam on Subaru, as well as near-IR bands Y, J, H, and Ks from VIRCAM on the VISTA telescope. All ALPINE galaxies except one are observed in ACS/HST F814W \citep{Scoville2007a, Koekemoer2007}. 
In addition, the galaxies are covered by the IRAC/Spitzer four channels from 3.6 $\mu$m to 8.0 $\mu$m from the SPLASH survey \citep{Capak2012, Steinhardt2014, Laigle2016} and MIPS/Spitzer 24 $\mu$m from the S-COSMOS \citep{Sanders2007, LeFloch2009}. 

Nine ALPINE galaxies are covered by NB2071 and NB2083 narrow-band imaging taken with the Multi-Object Infrared Camera and Spectrograph (MOIRCS; \citealp{Ichikawa2006, Suzuki2008}) on the Subaru Telescope, as part of the Charting Cluster Construction with VUDS \citep{LeFevre2015} and ORELSE \citep{Lubin2009} survey (C3VO, \citealp{Lemaux2020, Shen2021}). These observations are designed to target the [O \textsc{ii}] emission in the massive proto-cluster PCl J1001+0220 at $z\sim4.57$ and its surroundings in the COSMOS field \citep{Lemaux2018}. For more details on data reduction see Vanderhoof et al. (in prep). 

\subsection{X-ray Data} \label{sec:xray}

The Chandra COSMOS-Legacy Survey \citep{Civano2016, Marchesi2016a} identified 4016 X-ray sources down to a flux limit of $f_X \sim 8.9 \times 10^{-16}$ erg s$^{-1}$ cm$^{-2}$ in the 0.5-10 keV band. The Chandra COSMOS-Legacy catalog was matched with the UltraVISTA catalog using the Likelihood Ratio technique \citep{Sutherland1992}, which provide more statistically accurate result than a simple positional match (see more in \citealp{Laigle2016}). 
One ALPINE galaxy (``DEIMOS\_COSMOS\_845652'') was found to be an AGN detected with $L_{2-10\mathrm{keV}} = 10^{44.4}$ erg s$^{-1}$, using flux measured in the 0.5–2 keV band (corresponding to rest-frame 2–10 keV band at z $\sim$5) and assuming a $\Gamma$= 1.4, following eq.\ 4 in \citet{Marchesi2016c}. Such luminosity places it in the AGN regime, with $L_{2-10\mathrm{keV}} = 10^{42}$ erg\ s$^{-1}$ being the typical threshold to separate AGN and SFGs \citep{Marchesi2016b}. Nevertheless, none of our results would be affected by this galaxy, since all of our results are based on median stacking which eliminate the possible effect of single galaxy.

\subsection{Spectra} \label{sec:spec}

The rest-frame UV spectroscopic data from which the ALPINE sample is selected combine various large surveys. Out of the 105 ALPINE galaxies in the COSMOS field, 84 spectra are obtained from DEIMOS at the Keck in Hawaii \citep{Capak2004, Mallery2012, Hasinger2018}, and the remaining spectra are obtained from the VIMOS at the VLT in Chile \citep{LeFevre2005, Tasca2017}. 
The spectral resolution of them are different with R $\sim$ 230 for VIMOS and R $\sim$ 3200 for DEIMOS. See \citet{Faisst2020} for more details on the spectra of ALPINE galaxies. 

\subsection{Sample Selection}

We select the ALPINE galaxies in the COSMOS field that have [C \textsc{ii}] detections at S/N $\ge$ 3.5 \citep{Bethermin2020}. 
The [C \textsc{ii}] detection criterion selects galaxies that are more likely to have usable radio and FIR fluxes, since the presence of [C \textsc{ii}] both guarantees the redshift to be correct and increases the chance of appreciable FIR flux. Even so, this criterion potentially introduces selection bias. Both the median of stellar mass and that of SFR are increased 0.1 dex relative to the full APLINE sample when imposing the [C \textsc{ii}] detection cut. 
In addition, this criterion might potentially bias to galaxies having AGN activity. However, we applied various tests to our sample and do not find evidence of AGN activity, as discussed in Section \ref{sec:AGN}. 
In the 118 total ALPINE galaxies, the [C \textsc{ii}] line is detected in 75 of them. 
We select galaxies only in the COSMOS field, due to the availability of the 3 GHz observation. 
Our final sample contains 66 ALPINE galaxies, including 43 galaxies at $z\sim 4.5$ and 23 galaxies $z\sim 5.6$. 
We name the galaxies in the full, lower redshift, and higher redshift samples as ``CII-detected-all'', ``CII-detected-lz'', and ``CII-detected-hz'', respectively. 
The median properties of each sub-sample are summarized in Table \ref{tab:properties}.

\begin{deluxetable*}{c|cccccccc}
\tablecaption{Properties of three sub-samples \label{tab:properties}}
\tablewidth{0pt}
\tablehead{
\colhead{Sub-sample} & \colhead{Num. of } & \colhead{$\langle z\rangle$} & \colhead{$\langle$log(M$_*)\rangle$} & \colhead{$\langle$SFR$_{\mathrm{SED}}\rangle$} & \colhead{$\langle$SFR$_{\mathrm{UV}}\rangle$} & \colhead{$\langle$SFR$_{\mathrm{IR}}\rangle$} & \colhead{$\langle$SFR$_{\mathrm{tot}}\rangle$} \vspace{-2mm}\\
& \colhead{galaxies} & \colhead{}  & \colhead{log(M$_\odot$)} & \colhead{M$_\odot$ yr$^{-1}$} & \colhead{M$_\odot$ yr$^{-1}$} & \colhead{M$_\odot$ yr$^{-1}$} & \colhead{M$_\odot$ yr$^{-1}$}  \vspace{-1mm}\\
(1)& (2) & (3) & (4) & (5)  & (6) & (7) & (8)}
\startdata
 CII-detected-all & 66  &  4.57 & 9.8 & 29.4 & 16.9 & 14.7 & 31.6 \\
 CII-detected-lz  & 43  &  4.54 & 9.8 & 31.9 & 16.6 & 24.6 & 41.2 \\
 CII-detected-hz  & 23  &  5.63 & 9.8 & 26.9 & 17.3 & 11.6 & 28.9 \\
\enddata
\tablecomments{Column (1): Name of sub-samples. Column (2): The number of galaxies in each sub-sample. Column (3): The median of spectroscopic redshift derived from [C \textsc{ii}]. Column (4)-(5): The median of stellar mass and SFR$_{\mathrm{SED}}$ for each sub-sample derived from \textsc{Le Phare} (see \citealp{Faisst2020}). Column (6): The median of SFR$_{\mathrm{UV}}$ using the absolute restframe far-UV (FUV) luminosity provided by the ancillary ALPINE catalog \citep{Faisst2020} and calculated by SFR$_{\mathrm{UV}} = k_{UV} L_{FUV}$ with $k_{UV} = 1.47 \times 10^{-10}$ M$\odot$ yr$^{-1}$ L$\odot^{-1}$. Column (7): The median of SFR$_{\mathrm{FIR}}$ using $L_{IR}$ measured from the median-stacked FIR image and following the method described in Section \ref{sec:qTIR} (also see \citealp{Bethermin2020}). SFR$_{\mathrm{FIR}}$ is calculated by SFR$_{\mathrm{IR}} = k_{IR} L_{IR}$, where $k_{IR} = 1.02 \times 10^{-10}$ M$\odot$ yr$^{-1}$ L$\odot^{-1}$. Column (8): The median of SFR$_{\mathrm{tot}}$ by adding column (6) and (7).}
\end{deluxetable*}

\section{Methods} \label{sec:methods}

\subsection{Imaging Stacking and Flux Measurements} \label{sec:stacking}

Since none of the ALPINE galaxies are individually detected in the 3 GHz and 1.4GHz images, we carry out a stacking analysis. In order to systematically compare radio luminosity with IR luminosity and perform SED fitting, we stack FIR continuum maps from ALMA and available optical/NIR/MIR imaging. 
Two stacking methods that are commonly used are mean or median. The former method is preferred for a population that exhibits an approximately Gaussian flux distribution and requires the absence of close neighbors that contribute appreciable flux. On the other hand, median stacking is insensitive to the presence of outliers and has the advantage that all data can be used without qualification on neighboring objects (e.g., \citealp{Carilli2008, Karim2011, Man2016, Leslie2020}), though, it builds signal-to-noise more slowly. 
As our data do not present the ideal conditions for mean stacking, we adopt the median stacking method to all available bands.

We require accurate source positions in order to stack images at the locations of the galaxies. 
\citet{Faisst2020} measured the astrometric offset of individual ALPINE galaxies between the COSMOS2015 catalog and the HST images that aligned to the Gaia reference frame. We adopt these Gaia-corrected coordinates as the location of the galaxies for all images except the 3 GHz image. 
A small systematic astrometric offset has been identified by a positional matching of the 3 GHz sources with the COSMOS2015 catalog using a search radius of 0.8\arcsec \citep{Smolcic2017a}. We use the optical position in COSMOS2015 after correcting for the systematic offset of 3 GHz data using the best-fitting linear relations reported in \citet{Smolcic2017a}\footnote{
\begin{equation*}
\begin{aligned}
\rm{R.A. = R.A._{L16} + (-0.041R.A._{L16} + 6.1) / 3600} \\
\rm{decl. = decl._{L16} + (0.058decl._{L16} - 0.147)/3600}
\end{aligned}
\end{equation*} }. 
Note that no astrometric offset has been found in the 1.4 GHz data. The relative and absolute astrometry of the 1.4 GHz image are 130 and $<$ 55 mas, respectively \citep{Schinnerer2007}. The relative astrometry is tested by comparing the position of sources extracted from each single pointing and that from the combined mosaic, and with the VLA FIRST survey catalog \citep{Becker1995}. 

The stacking and detection are straightforward. 
For galaxies in each sub-sample and each band, we create a 40\arcsec\ $\times$ 40\arcsec\ cutout of image, centered on the applied coordinate of each galaxy. We then calculate the median stack image of N$_\mathrm{objs}$. 
To create stacked images free from projected bright sources, we remove sources detected above $>5\sigma$ significance based on the 3 GHz catalog \citep{Smolcic2017a}, 1.4 GHz catalog \citep{Schinnerer2010}, and serendipitous detection catalog from \citet{Bethermin2020} for FIR.  
These sources are removed by fitting a two-dimension elliptical Gaussian within a 4\arcsec$\times$ 4\arcsec\ box using the astropy fitting tools \citep{Astropy2013, Astropy2018}. These sources are found to be point sources in all radio and FIR images by visual examination. This box size is then chosen based on the beam size of radio/FIR images and to be slightly larger than these point sources. 
Except one of 3 GHz source that is found to be 1.17\arcsec\ away from an ALPINE galaxy (DEIMOS\_COSMOS\_842313), the separations between these point sources and our sample are larger than the beam size of the corresponding image, which excludes the case that fluxes in the beam area of ALPINE galaxies are removed.

For radio and FIR stacks, we adopt the CASA task imfit to measure the peak and total fluxes by fitting one Gaussian component restricted to the central 3\arcsec\ $\times$ 3\arcsec\ box. The detection threshold is set to a peak above three times the local rms noise. For non-detections, we adopt the 3$\sigma$ as the upper limit, where $\sigma$ is the rms of the stack image obtained by CASA task imstat derived in the full 40\arcsec\ $\times$ 40\arcsec\ stack image. 
For optical/NIR/MIR stacks, we adopt the SExtractor \citep{Bertin1996} and measure the 3\arcsec\ aperture flux at the center of each stack image. The detection threshold is set to the aperture flux above five times the associated aperture flux error. For non-detections, we adopt the 3$\sigma$ as the upper limit, where $\sigma$ is measured from the flux scatter of randomly distributed 3\arcsec\ apertures. An additional flux correction is applied to IRAC images to obtain total fluxes following \citet{Ilbert2008}. We show all of the stacked images for the CII-detected-all sample in Figure \ref{fig:allstacks} in the Appendix. 

To estimate the uncertainties on the stacked fluxes that represent the sample variance, we perform a bootstrap analysis by randomly drawing, with replacement, N$_\mathrm{objs}$ galaxy cutouts (from our initial list of galaxies) and creating a new median stack. We then measure the fluxes for each bootstrap stack using the same method as was done for the original stack.  We repeat this process 100 times. When CASA imfit failed, such that no Gaussian component is successfully fitted, we adopt the 3\arcsec\ aperture flux. It has been found that fluxes obtained by 2D Gaussian fitting and aperture measurements are overall in excellent agreement \citep{Bethermin2020}. We adopt the 16th and 84th percentiles of bootstrap fluxes as our errors on the peak and total fluxes, to represent the sample variance. Note that we do not apply detection threshold for the bootstrap fluxes.

Using ALPINE data, \citet{Romano2021} found that a large fraction ($\sim$40\%) of [C \textsc{ii}]-detected ALPINE galaxies are mergers. The majority of them have small projected distances between the centers of the merger components ($r_p \lesssim 10$ kpc) as found by inspecting their [C \textsc{ii}] intensity map. Thus, we expect the majority of components in these mergers are covered by the 3\arcsec\ box (corresponding to 19 kpc at z = 5) for radio/FIR images or the 3\arcsec\ aperture for optical/NIR images. The widespread presence of such merging activity may introduce smearing and offsets in the radio images, which might be one of the reasons that we observe offsets in our radio stacked images (see Section \ref{sec:radio_detection}).

\subsection{Estimating the AGN contribution and SFR from SED fitting} \label{sec:sed}

We employed the SED fitting Code Investigating GALaxy Emission (\textsc{cigale}) \citep{Boquien2019} in order to constrain the AGN contribution to the IR luminosity of our stacked samples in a self-consistent framework that considers the energy balance between the UV/optical and IR. 
We set the templates and parameters in \textsc{cigale} to be consistent with previous SED fitting for ALPINE galaxies that was done with \textsc{Le Phare} \citep{Arnouts1999, Ilbert2006} as described in \cite{Faisst2020}. 
In detail, we adopted a delayed exponential star formation history (SFH, sfhdelayed) allowing the range of $\tau$ and age to be similar to those used in \cite{Faisst2020}. 
We assumed a \citet{Chabrier2003} IMF and the stellar population synthesis models presented by \citet{Bruzual2003} with solar (Z$_\odot$) and 0.2 Z$_\odot$ metallicity. 
Dust attenuation follows \citet{Calzetti2000}'s extinction law allowing colour excess $E_s$(B-V) to vary from 0 to 0.5. 
Note that the absorption UV bump feature produced by dust at 2175 \AA\ is set to zero because the stacked SEDs do not have enough spectral resolution to meaningfully constrain it. 
For the dust emission module, we adopted the dust templates of \citet{Draine2014} to keep consistent with the best-fit IR SED template found for ALPINE analogs in the COSMOS field, which was used to derive the IR luminosity from the rest-frame 158$\mu$m continuum fluxes \citep{Bethermin2020}. 
This best-fit IR SED template has a mean intensity of the radiation field $\langle U \rangle = 50$ with fixed values of the maximal radiation field ($U_{\mathrm{max}} = 10^6$), the fraction illuminated from $U_{\mathrm{min}}$ to $U_{\mathrm{max}}$ ($\gamma$ = 0.02), the power law index ($\alpha=2$), and the polycyclic aromatic hydrocarbon (PAH = 2.5\%) \citep{Bethermin2015, Bethermin2017, Magdis2012}. 
Following this best-fit IR SED template, we set parameters in the dust model, only allowing the the minimal radiation field $U_{\mathrm{min}}$ to vary from 30 to 40, corresponding to $\langle U \rangle = 37 - 50$.  

For the AGN module, we adopt the latest SKIRTOR template that is implemented in \textsc{cigale} \citep{Yang2020}. 
We retain the default parameters in the AGN module, other than setting the viewing angle $i$ to 30$^{\circ}$ and 70$^{\circ}$ for type 1 and type 2 AGNs, respectively, and a full range of AGN fraction ($f_{\mathrm{AGN}}$) from 0 to 0.9, a fraction that denotes the contribution from the AGN to the total IR luminosity \citep{Ciesla2015}. 
We note that only two viewing angles are selected in order to obtain the most informative parameters. \citet{RamosPadilla2022} have tested on the effect of the viewing angles in AGN classification and found that limiting viewing angles representing these two types of AGNs (30$^{\circ}$/70$^{\circ}$) does not affect the estimated physical parameters (i.e. $f_{\mathrm{AGN}}$) as compared to using the full range of viewing angles. 

For radio synchrotron emission, either from star formation or AGN, we allow the $q_{\mathrm{TIR}}$ to vary in the range of 1.5 to 3.0, to cover the $q_{\mathrm{TIR}}$ of local normal star-forming galaxies \citep{Bell2003} and those expected for SFGs and/or AGNs at $z \sim 5$ \citep{Delhaize2017}.  
We adopt a single spectral slope of power-law synchrotron emission $\alpha_{radio}$ = 0.7, since we do not have detections in enough radio bands to constrain the radio spectral slope (see discussion for the selection of $\alpha$ in \ref{sec:radio_detection}). 

Finally, we adopt the `pdf analysis' method in \textsc{cigale} to compute the likelihood ($\chi^2$) for all the possible combinations of parameters and generate the marginalized probability distribution function (PDF) for each parameter and each galaxy stack. 
More details of the parameter settings are shown in Table \ref{tab:sed}.
In addition, we adopt a mock analysis in \textsc{cigale} to derive the uncertainty of estimated parameters based on the photometric errors. The mocks are built based on the best fit for each object and modied by sampling from a Gaussian distribution with the same standard deviation as the uncertainty on the observation. See more discussion of this method in the Appendix of \citealp{shen2020a}). 
We generate 100 mocks for each stacked SED and adopt the 16th/84th percentiles as the uncertainty of each parameter. 

We run \textsc{cigale} on photometry measured from the stacked ground-based observations in u, B, V, r$^{+}$, i$^{+}$, z$^{++}$, Y, Y$_{\mathrm{HSC}}$, J$_{\mathrm{HSC}}$, H$_{\mathrm{HSC}}$, H$_{\mathrm{HSC}}$, Ks, the intermediate-bands IA427, IA464, IA484, IA505, IA527, IA574, IA624, IA679, IA709, IA738, IA767, and IA827, as well as from the space-based observation in F814W/HST, all four IRAC/Spitzer channels and MIPS/Spitzer at 24$\mu$m.
The best-fitted model of three stacked SEDs are shown in Figure \ref{fig:sed}. The reduced $\chi^2$ of the best-fitted SEDs are 1.5, 0.71, and 0.70 for the CII-detected-all, CII-detected-lz, and CII-detected-hz stacks, respectively. We note that the redshift of CII-detected-all is set to its median spectroscopic redshift $z = 4.57$, however the stacked SED includes galaxies from $z = 5.5$, which might be the reason for the larger reduced $\chi^2$. 
We note that, with our current photometry data set, we are not able to constrain the AGN and dust emission at the same time. Therefore, we only allow a small variation on the dust module to keep consistent with the best-fit IR SED template from \cite{Bethermin2020}, so that CIGALE fitting was performed mainly to estimate two parameters ($f_{\mathrm{AGN}}$ and viewing angle) associated with AGN module. The former is defined as the AGN contribution to IR luminosity ($L_{\mathrm{AGN}} = f_{\mathrm{AGN}}\times L_{\mathrm{IR}}$; \citealp{Ciesla2015}). The later separates the typical AGN templates for Type-1 and Type-2 AGN. 
As a sanity check, we compare the stellar mass and SFR from the stacked SEDs using \textsc{cigale} and the median of them from the ancillary ALPINE catalog derived from \textsc{Le Phare}. 
Though we observe some differences in the recovered parameters from the stacked \textsc{cigale} fitting and the median parameters of the \textsc{Le Phare} fitting for various stacked samples, none of the differences are statistically significant.

\begin{deluxetable}{c|c}
\tablecaption{Parameter ranges used in the SED fitting with \textsc{cigale}. \label{tab:sed}}
\tablewidth{0pt}
\tablehead{
\colhead{Parameter} & \colhead{Values}}
\startdata
\multicolumn{2}{c}{SFH sfhdelayed}  \\
\hline
$\tau$ [Myr]  &  100, 500, 1000, 3000  \\
Age [Myr]  & 50 - 1100 \\
\hline
\multicolumn{2}{c}{SSP \citep{Bruzual2003}} \\
\hline
IMF & \citet{Chabrier2003} \\
Metallicity($Z_\odot$) &  1, 0.02 \\ 
\hline
\multicolumn{2}{c}{Dust Attenuation \citep{Calzetti2000}}  \\
\hline
E(B-V)$_l$ & 0.0 - 1.1 \\ 
E(B-V)$_{factor}$ & 0.44 \\
Slope of the power law & -0.5, -0.25, 0 \\ 
\hline
\multicolumn{2}{c}{Dust emission \citep{Draine2014}} \\
\hline
Mass fraction of PAH & 2.5\% \\
Minimum radiation field & 30, 35, 40 \\
Power slope $dU/dM \propto U^{-\alpha}$ & 2.0 \\
Dust fraction in PDRs & 0.02 \\
\hline
\multicolumn{2}{c}{AGN emission } \\
\hline
$\theta$ & 30, 70 \\
$f_{\mathrm{AGN}}$ & 0 - 0.9 \\
\hline
\multicolumn{2}{c}{radio emission } \\
\hline
$q_{\mathrm{IR}}$ & 1.5, 1.75, 2.0, 2.25, 2.5, 3.0 \\
$\alpha$ & 0.7 
\enddata
\end{deluxetable}

\section{Results} \label{sec:result}

\subsection{Radio detection and luminosity} \label{sec:radio_detection}

\begin{figure*}
    \centering
    \includegraphics[width=\textwidth]{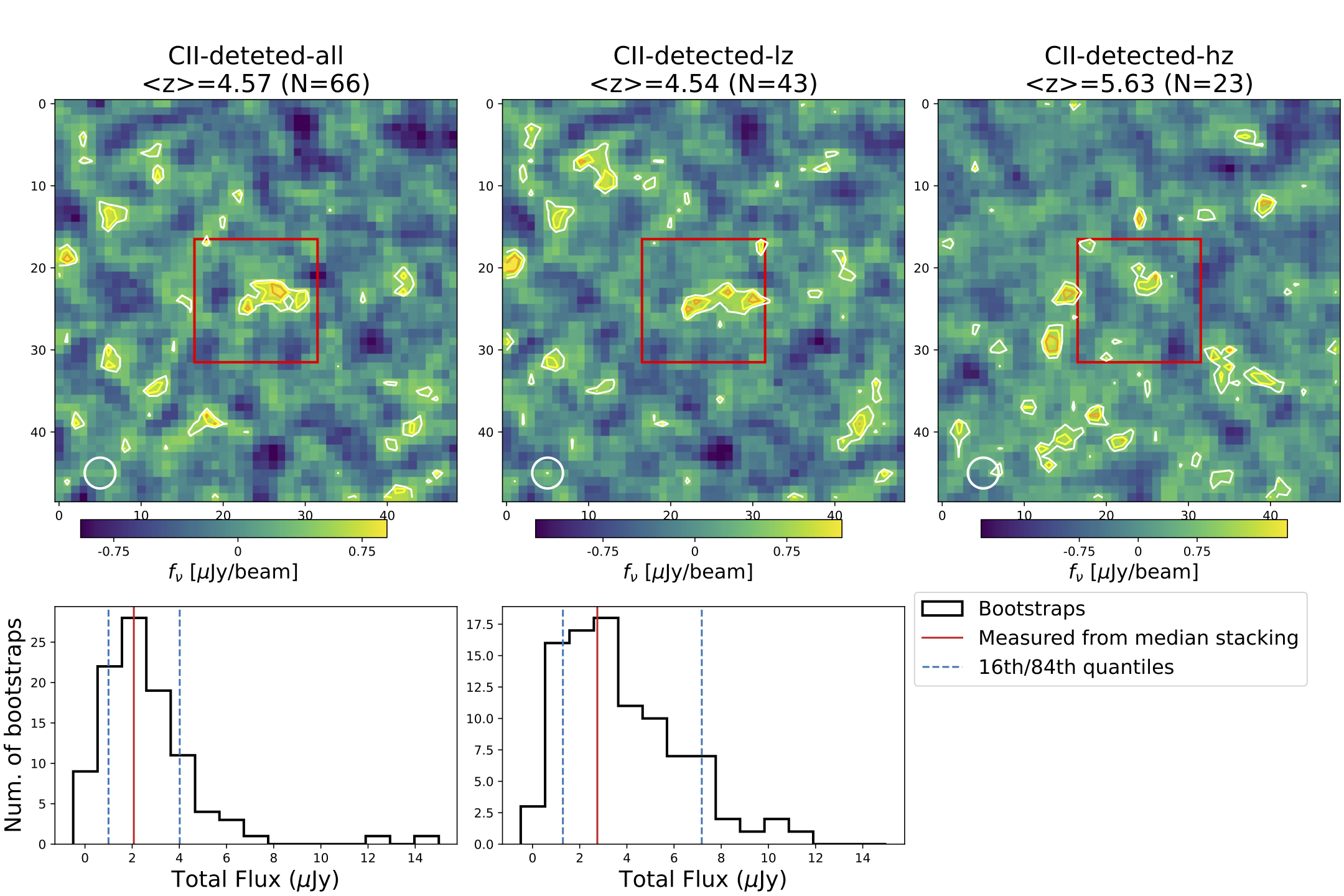}
    \caption{\textit{Top:} The $10\arcsec \times 10\arcsec$ median stacked 3 GHz image of CII-detected-all, CII-detected-lz and CII-detected-hz samples (see text for the meaning of these samples) from left to right. In each panel, black contours show 1.5, 2, and 2.5 $\times$ rms obtained by CASA task \textsc{imstat} in the full stack image. The central 3\arcsec $\times$ 3\arcsec\ box used for detection is marked in red, and the 0.75\arcsec\ beam is shown in the bottom left corner. \textit{Bottom:} The corresponding histograms of bootstrap total fluxes obtained from 100 bootstrap stacks for the CII-detected-all and CII-detected-lz samples. In each panel, the measured total flux and 16th/84th percentiles of bootstrap fluxes are shown as red solid and green dotted lines, respectively. The histogram of the CII-detected-hz is not shown, since no detection is obtained in the original stacking and in the most of the bootstrap stacking realizations. }
    \label{fig:radio}
\end{figure*}

In Figure \ref{fig:radio}, we show the $10\arcsec \times 10\arcsec$ median-stacked 3 GHz image of CII-detected-all, CII-detected-lz and CII-detected-hz, from left to right, respectively, and the corresponding histograms of bootstrap total fluxes in the bottom panels. 
We successfully detect a signal in the stacked 3 GHz image of CII-detected-all and CII-detected-lz with their peak fluxes at $3.2\sigma$ and $3.9\sigma$, respectively, where $\sigma$ is the associated error on the peak fluxes. 
No significant detection ($>3\sigma$) is obtained in the stacked 3 GHz image of CII-detected-hz, which is most likely due to the smaller number of galaxies in the CII-detected-hz sample. To test this, we randomly select 23 galaxies from the CII-detected-all samples, the same number of galaxies in the CII-detected-hz sample, without replacement, and apply the same median stacking and flux measurement. We do not recover any significant detection in the majority (73\%) of the iterations. 
In addition, none of these sub-samples are significantly detected in their stacked 1.4 GHz image. 
Fluxes and uncertainties from bootstraps are shown in Table \ref{tab:radio_data}. 
We also test using mean stacking, which directly increases the S/N of an image as $\sqrt n$, where n is the number of galaxies. We obtain higher significant detections for CII-detected-all and CII-detected-lz with their peak fluxes at $4.7\sigma$ and $6\sigma$, respectively. 
We note that these increased S/N only for statistical argument, as they are likely to be contaminated by neighbors at much higher significance levels. 
In addition, we recover stacked 3GHz fluxes that are statistically identical using the position of [C \textsc{ii}] emission (see Table C.1 in \citet{Bethermin2020}), done without coordinate correction described in Section \ref{sec:stacking}. 
It should be noted that we do not detect signal in the stacked 3GHz image when including all ALPINE galaxies, which is most likely that the fainter sources dilute the signal in the stacks.

Though we use the peak flux for detection significance, the total fluxes are adopted because we see an elongated shape in our stacked image. 
The elongated shape might be due to various astrophysical and astrometric effects that cause an effective blurring of the stacked image. The astrophysical effects might be the dominant causes, due to the presence of mergers in our sample, the physical extend of SFGs and possible offset between optical and radio sources. In fact, it has previously been found that the physical extent of SFGs at 3 GHz is 1-2kpc out to z$\sim$2.25 \citep{Jimenez-Andrade2019}, which corresponds to 0.15\arcsec-0.3\arcsec at z=4.6. In addition, it could also be affected by sub-pixel variations between images. For example, a small (0.1\arcsec) astrometric offset persists between the optical catalog and the true 3 GHz source position after correcting for the systemic offset \citep{Smolcic2017a}. 
Therefore, we use the total flux densities as our final radio fluxes.

The measured 3 GHz flux densities ($S_{\mathrm{3GHz,total}}$) are converted to rest-frame 1.4 GHz luminosities via: 
\begin{equation}
    L_{\mathrm{1.4GHz}} = \frac{4\pi D^2_L}{(1+z)^{1+\alpha}} \left( \frac{\mathrm{1.4GHz}}{\mathrm{3GHz}} \right) ^\alpha S_{\mathrm{3GHz}}
\end{equation}
where $D_L$ is the luminosity distance to the galaxy, and $\alpha$ is the spectral index. 
Because all three sub-samples are not detected at 1.4 GHz, we calculate the lower limit on the spectral index, as listed in Table \ref{tab:radio_data}. 
\citet{Smolcic2017} cross-matched the 3 GHz catalog with the 1.4 GHz catalog in COSMOS with a maximum separation of 1.3\arcsec and obtained the spectral index of on average $\alpha = -0.73 \pm 0.35$. \citet{Smolcic2017b} showed that the spectral index tends to be steeper with increasing redshift. However, such a trend might not be robust at high redshift due to the small sample of galaxies at $z \ge 3$, flux limitation on these observations, and, at higher rest-frame frequency as increasing redshift, the spectral index becomes flatter due to the free-free emission dominating the radio emission. 
\citet{Delhaize2017} further selected SFGs in the combined 3 GHz and 1.4 GHz catalogs and quantified the median spectral index of SFGs to be -0.7 at $z < 2$ and -0.8 at $z > 2$. These two spectral index values are typical values used for radio sources.
This $\alpha$ value is consistent with that found in the sub-mJy radio population \citep{Ibar2009}, submillimetre galaxies at z$\sim$2 \citep{Ibar2010}, and the lower limits of $\alpha$ from the ALPINE sub-samples. 
Therefore, we adopt a simple $\alpha = -0.7$ in the calculation of rest-frame 1.4 GHz luminosity. 
We examine the possible effect on $q_{\mathrm{TIR}}$ using a steeper and flatter spectral index ($\alpha = -0.8$, $\alpha = -0.6$) in Section \ref{sec:discussion}. 

\begin{deluxetable*}{c|ccccccc}
\tablecaption{Radio flux densities at 3GHz and 1.4GHz \label{tab:radio_data}}
\tablewidth{0pt}
\tablehead{
\colhead{Sub-sample} & \colhead{$S_{\mathrm{3GHz, peak}}$} & \colhead{$S_{\mathrm{3GHz,total}}$} & \colhead{rms$_{\mathrm{3GHz}}$} & \colhead{rms$_{\mathrm{1.4GHz}}$} & \colhead{$\alpha_{\mathrm{1.4-3 GHz}}$} & \colhead{$L_{\mathrm{1.4GHz}}$} \vspace{-2mm}\\
& \colhead{$\mu$Jy/beam} & \colhead{$\mu$Jy}  & \colhead{$\mu$Jy/beam} & \colhead{$\mu$Jy/beam} & & \colhead{$10^{24}$ W Hz$^{-1}$} \vspace{-1mm}\\
(1)& (2) & (3) & (4) & (5)  & (6) & (7)}
\startdata
 CII-detected-all &  $1.08_{-0.21}^{+0.56}$  & $2.07_{-1.10}^{+1.92}$ & 0.36 & 1.94 & $\geq$-1.35 & $0.44_{-0.24}^{+0.41}$ \\
 CII-detected-lz  &  $1.21_{-0.20}^{+0.68}$  & $2.75_{-2.02}^{+3.87}$ & 0.44 & 2.38 & $\geq$-1.25 & $0.58_{-0.43}^{+0.82}$ \\
 CII-detected-hz  &  -                       & -                      & 0.61 & 3.02 & $\geq$-2.11 & $< 0.60$  \\
\enddata
\tablecomments{Column (1): Name of sub-samples. Column (2) and (3) peak and total fluxes measured in the 3 GHz stack images. The associated error are the 16th and 84th percentiles of bootstrap fluxes. Column (4) and (5) the rms of the 3 GHz and 1.4 GHz stack images. (6) 3$\sigma$ lower limit of spectra index. (7) rest-frame 1.4 GHz luminosity for CII-detected-all/-lz and 3$\sigma$ upper limit for CII-detected-hz. }
\end{deluxetable*}

Radio luminosity can be used as an SFR indicator for SFGs. We estimated radio derived SFR (SFR$_{\mathrm{radio}}$) using the SFR formula from 1.4 GHz from \citet{Bell2003} and converting Salpeter IMF to Chabrier IMF by multiplying by a factor of 0.6. 
The SFR$_{\mathrm{radio}}$ are $147^{+136}_{-78}$, $193^{+271}_{-142}$, $<200$ M$_\odot$ yr$^{-1}$ for CII-detected-all, -lz, and -hz, respectively. The uncertainties are derived from the average of 16th/84th percentiles of bootstrap radio fluxes. 
These values are considerably larger than the average total SFR measured combining ALPINE continuum stacking and ancillary UV data, as shown in Table \ref{tab:properties}. 
Specifically, using the stacked FIR fluxes and the median far-UV luminosity, L$_{\mathrm{FUV}}$, provided by the ancillary ALPINE catalog \citep{Faisst2020}, and following equation 9 in \citet{Bethermin2020}, we calculate the full SFR (SFR$_{\mathrm{tot}}$) of $32\pm14$, $41\pm18$, and $29\pm21$ M$_\odot$ yr$^{-1}$ for CII-detected-all, -lz, and -hz, respectively. The uncertainties are the average of 16th/84th percentiles of bootstrap FIR fluxes and 16th/84th percentiles of L$_{\mathrm{FUV}}$.  
Although, the difference between radio- and UV+IR-derived SFR are not statistically significant, due mostly to the extremely large errors of the former estimate, such a difference may suggest that either an appreciable AGN component exists that contributes to the radio luminosity or the SFR$_{\mathrm{radio}}$ formula, which is calibrated based on the local IR-radio correlation, might not be appropriate for these galaxies or some combination there of. 
Therefore, we further investigate the IR-radio correlation of our sample as compared to other studies in the literature in section \ref{sec:qTIR} and the possible AGN contamination in section \ref{sec:AGN}.

\subsection{The IR-Radio correlation} \label{sec:qTIR}
The IR-Radio correlation can be quantified by the parameter $q_\mathrm{{TIR}}$, defined as the logarithmic ratio of the total infrared (8 - 1000 $\mu$m) and 1.4 GHz luminosities:
\begin{equation} \label{eq:qTIR}
    q_{\mathrm{TIR}} = log\left(\frac{L_\mathrm{{TIR}}/\mathrm{W}}{3.75 \times 10^{12} \mathrm{Hz}}\right) - log\left(\frac{L_\mathrm{{1.4GHz}}}{\mathrm{W\ Hz^{-1}}}\right) 
\end{equation}
where $L_\mathrm{{TIR}}$ is the rest-frame total infrared (8 - 1000 $\mu$m) luminosity, and $L_\mathrm{{1.4GHz}}$ is the rest-frame radio luminosity at 1.4 GHz \citep{Condon1992, Bell2003}. We use the same method as \citet{Bethermin2020} to convert the rest-frame 158 $\mu$m stack flux to $L_\mathrm{{TIR}}$. \citet{Bethermin2020} identified the best-fitted IR SED template by fitting to a mean stacked continuum FIR SED for ALPINE galaxy analogs in terms of redshift, stellar mass and SFR in the COSMOS field. 

\begin{figure*}
    \centering
    \includegraphics[width=\textwidth]{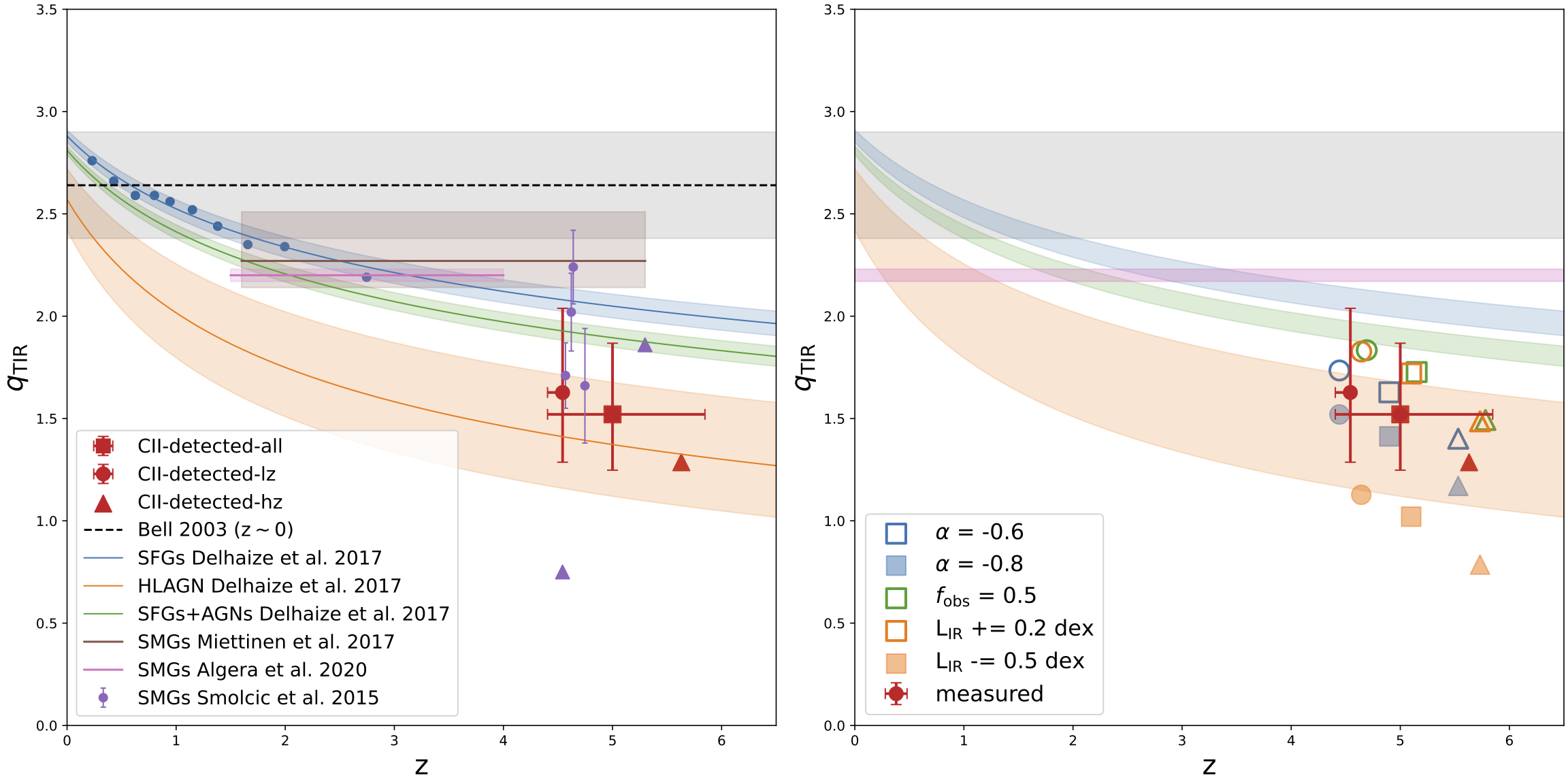}
    \caption{\textit{Left:} The stacked $q_{\mathrm{TIR}}$ versus median redshift for the three ALPINE subsamples as compared to literature. The measured $q_{\mathrm{TIR}}$ of CII-detected-all/-lz are shown as the red dot and square with error bars. The 3$\sigma$ lower limit of CII-detected-hz is shown as triangle. The $q_{\mathrm{TIR}}$ errors includes the spread of of IR and radio fluxes obtained from bootstrapping. The redshift error bars represent the redshift range in each sub-sample. The blue dots are the median $q_{\mathrm{TIR}}$ for SFGs binned by redshift adopted from \citet{Delhaize2017}. The blue solid line are the power-law fit and its error shown as shaded region. The green and orange lines and shaded regions are the best power-law fits and associated errors for all galaxies (labeled as "SFGs+AGNs") and for only HLAGN \citep{Delhaize2017}. The  black dotted line and shaded region are the local value of \citet{Bell2003} ($q_{\mathrm{TIR}} (z\sim0) = 2.64 \pm 0.02$) and associated spread of 0.26, respectively. The brown and pink lines show the median $q_\mathrm{{TIR}}$ measured for SMGs at $z = 1.6-5.3$ \citep{Miettinen2017} and at $z = 1.5-4$ \citep{Algera2020}, respectively. The purple markers are adopted from \citet{Smolcic2015}, with triangle represent the lower limits. \textit{Right:} The fiducial $q_{\mathrm{TIR}}$ for the three ALPINE subsamples. The measured/lower limit $q_{\mathrm{TIR}}$ are shown as red solid markers with error bars. The fiducial $q_{\mathrm{TIR}}$ are shown as: blue open markers for using $\alpha = -0.6$, blue shaded markers for using $\alpha = -0.8$, orange open markers for increasing the $L_{\mathrm{IR}}$ by 0.2 dex due to different IR SED templates, orange shaded markers for decreasing the $L_{\mathrm{IR}}$ by 0.5 dex due to different IR SED templates, and green open markers for apply a correction on the obscured fraction of star formation (See section \ref{sec:discussion}). The points are offset in redshift for clarity. The shaded regions are the same as those shown in the right panel for direct comparison. } 
    \label{fig:qTIR}
\end{figure*}

The $q_{\mathrm{TIR}}$ are $1.52_{-0.27}^{+0.35}$ and $1.63_{-0.34}^{+0.41}$ for CII-detected-all and -lz, respectively, and lower limit of 1.29 for CII-detected-hz. The lower/upper uncertainties of $q_{\mathrm{TIR}}$ are the difference between 16th/84th percentiles and median of $q_{\mathrm{TIR}}$ distribution as calculated from IR and radio bootstrapping fluxes, which represents the sample variance. 
Note that the systematic uncertainty of assuming a single FIR SED is not accounted for here (this is done in Section \ref{sec:discussion}). 
Figure \ref{fig:qTIR} shows the stacked $q_{\mathrm{TIR}}$ and associated uncertainties versus the median redshift of the three ALPINE sub-samples. For clarity, the redshift of CII-detected-all is offset from its median value (z=4.57). The errorbar on the x-axis represents the redshift range of each sub-sample. 
Note that we do not consider AGN in the calculation of $q_\mathrm{{TIR}}$. Although, the data are not sufficient to confirm whether AGN contribute to the radio band either through spectral index or SFR estimates, we do not find evidence of dominant AGN activity in the optical-to-far-IR SED, X-ray and UV spectra. See further discussion related to AGN in Section \ref{sec:fAGN}.

We compare our results of $q_{\mathrm{TIR}}$ to several other studies in the literature.
Firstly, a $q_{\mathrm{TIR}}$ $= 2.64 \pm 0.02$ with a scatter of 0.26 dex was measured by \citet{Bell2003} in 162 local normal SFGs . 
Our measured $q_\mathrm{{TIR}}$ are clearly offset from the $q_{\mathrm{TIR}}$ of local SFGs, even considering their large errorbars. 
More specifically, the $q_{\mathrm{TIR}}$ of the CII-detected-all and CII-detected-lz sub-samples are offset from the local $q_{\mathrm{TIR}}$ value by 3.2 and 2.5 $\sigma$, where $\sigma$ includes the uncertainties on $q_{\mathrm{TIR}}$ of ALPINE subsamples and that of local SFGs. 
More recently, \citet{Delhaize2017} quantified $q_{\mathrm{TIR}}$ out to $z\sim 3$ using the same 3 GHz image as this paper and Herschel data \citep{Lutz2011, Oliver2012} in the COSMOS field. They have employed a doubly-censored survival analysis to include both lower and upper limits and quantified the evolution of $q_{\mathrm{TIR}}$ for SFGs, all galaxies including SFGs and AGNs, and only moderate-to-high radiative luminosity AGNs (HLAGN) samples. 
The $q_\mathrm{{TIR}}$ of ALPINE sub-samples are consistent with this trend, though our values tend towards the lower end of $q_\mathrm{{TIR}}$ range of models where some AGN contribution exists.

It has also been found that the $q_\mathrm{{TIR}}$ of SMGs at high redshift ($z > 1.5$) are in general lower than local values for normal SFGs \citep{Murphy2009, Michalowski2010, Thomson2014, Smolcic2015, Miettinen2017, Algera2020}. 
\citet{Miettinen2017} studied the physical properties of 16 SMGs in the redshift range of $z = 1.6-5.3$ in the COSMOS field. They found a median $q_\mathrm{{TIR}}$ of $2.27^{+0.27}_{-0.13}$, shown as the brown line with the 16th–84th quantiles as shaded region, using radio luminosity derived from 325 MHz (corresponding to the rest-frame 1.4 GHz at z = 3.3) and IR luminosity derived from SED fitting by the MAGPHYS code. 
Recently, \citet{Algera2020} measured $q_\mathrm{{TIR}}$ of 273 SMGs that have 1.4 GHz detections, and found a median $q_\mathrm{{TIR}} = 2.20 \pm 0.03 $ independent of redshift, shown as the pink line and shaded region. 
In addition, \citet{Smolcic2015} measured the $q_\mathrm{{TIR}}$ of six spectroscopically-confirmed SMGs at $z > 4$ in the COSMOS field. Using survival analysis, they found a mean and standard derivative of $q_\mathrm{{TIR}} = 1.95 \pm 0.2$. The $q_\mathrm{{TIR}}$ of \citet{Smolcic2015} are shown as purple dots with errorbars and their upper limits are shown as triangles. 
We see our stacked $q_\mathrm{{TIR}}$ are consistent with values from \citet{Smolcic2015} and broadly in line with the trend of \citet{Miettinen2017} and \citet{Algera2020} within the errors. 
However, we point out that the physical properties of these SMGs, such as stellar mass and star-formation rate, are extremely different from galaxies in ALPINE. 
The average stellar mass and the range of star formation rate of SMGs in \citet{Smolcic2015} is $1.4\times10^{11} M_\odot$ and 600 - 2000  M$_\odot$ yr$^{-1}$, respectively, both well in excess of those values of the ALPINE galaxies studied here. 
That our stacked $q_\mathrm{{TIR}}$ is lower than $q_\mathrm{{TIR}}$ values for normal SFGs potentially indicates an AGN contribution and/or that lower $q_\mathrm{{TIR}}$ values are a general property of high-redshift ($z > 4$) SFGs.

\subsection{Possible AGN Contribution} \label{sec:AGN}

While powerful broad-line AGN were selected against when constructing the ALPINE sample, the prevalence of AGN was previously unknown. To address possible AGN contribution to the $q_\mathrm{{TIR}}$, we apply three approaches: the SED fitting using the stacked multi-wavelength photometry, stacking X-ray data and rest-frame UV spectra.

\subsubsection{AGN fraction}\label{sec:fAGN}

Firstly, we quantify the fraction of AGN contribution to IR luminosity using \textsc{cigale} for the stacked SED of the three ALPINE sub-samples
The best-fitted $f_{\mathrm{AGN}}$ are $0.00^{+0.18}$, $0.15^{+0.30}_{-0.15}$, and $0.18^{+0.52}_{-0.18}$ for CII-detected-all/-lz/-hz sub-samples, where the uncertainty of $f_{\mathrm{AGN}}$ are the 16th/84th percentiles of 100 mocks, regardless of AGN type. 
The stacked SED of CII-detected-lz and CII-detected-hz are best-fitted to Type-2 AGN with a viewing angle of 70$^\circ$ and Type-1 AGN with a viewing angle of 30$^\circ$, respectively. 
The corresponding bolometric $\mathrm{L_{AGN}}$ from the best-fitted models of CII-detected-lz and -hz sub-samples are both $10^{44.2}$ erg/s, though their $f_{\mathrm{AGN}}$ statistically consistent with no AGN activity. The upper limits of $\mathrm{L_{AGN}}$ from Type-1/-2 mocks are $10^{44.7}$/$10^{45.0}$ erg/s, $10^{44.7}$/$10^{45.5}$ erg/s and $10^{44.7}$/$10^{45.6}$ erg/s for CII-detected-all, -lz and -hz sub-samples, respectively. 

\begin{figure*}
    \centering
    \includegraphics[width=0.45\textwidth]{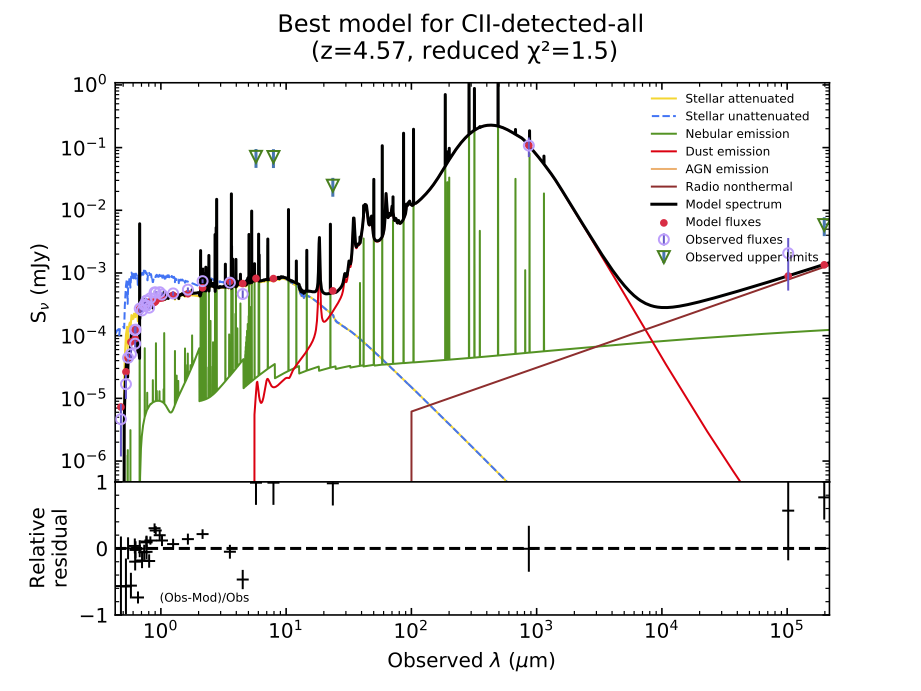} 
    \includegraphics[width=0.45\textwidth]{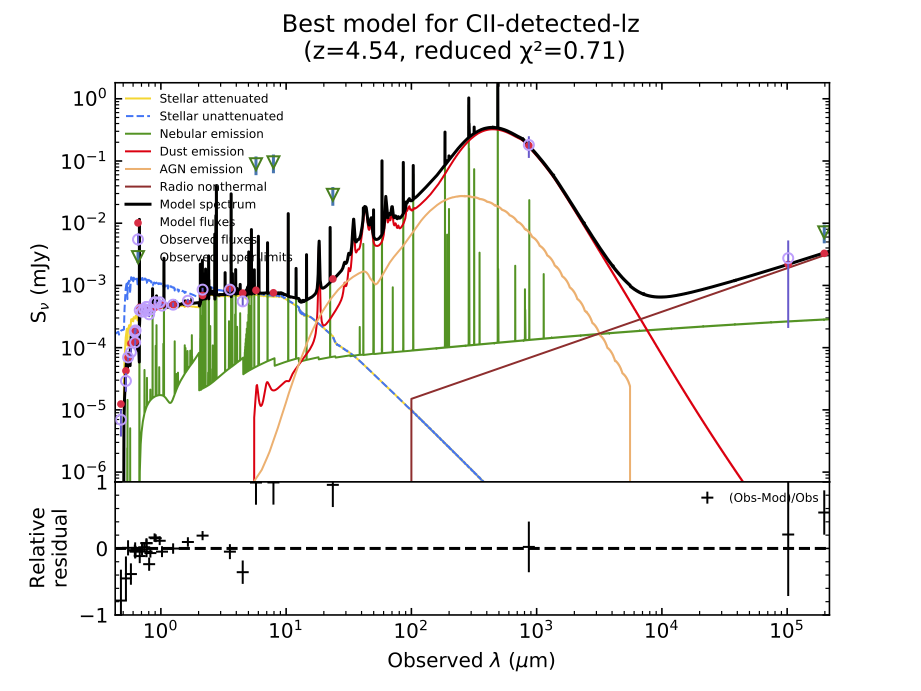} 
    \includegraphics[width=0.45\textwidth]{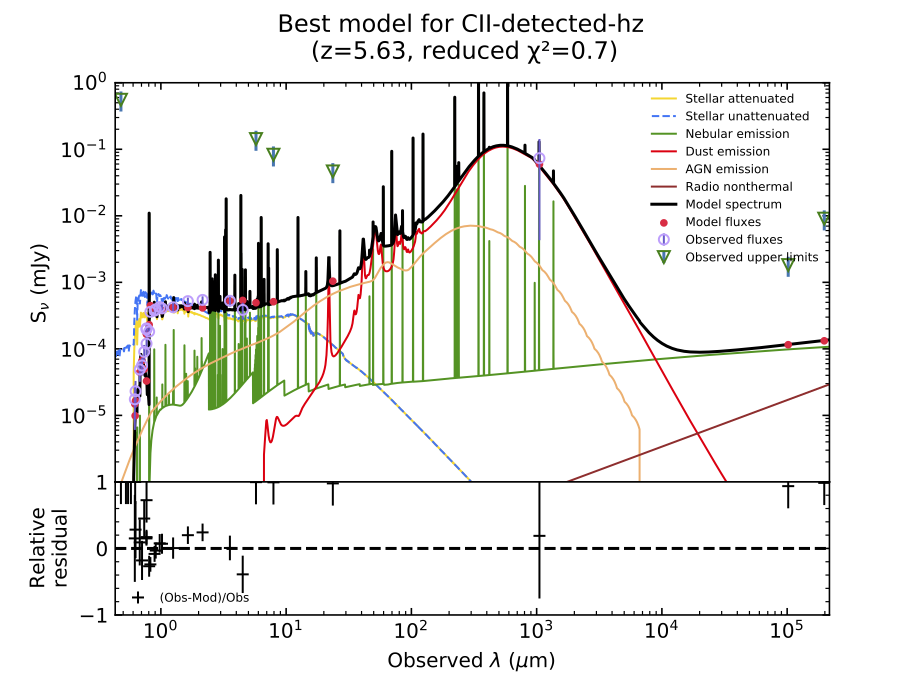}
    \includegraphics[width=0.45\textwidth]{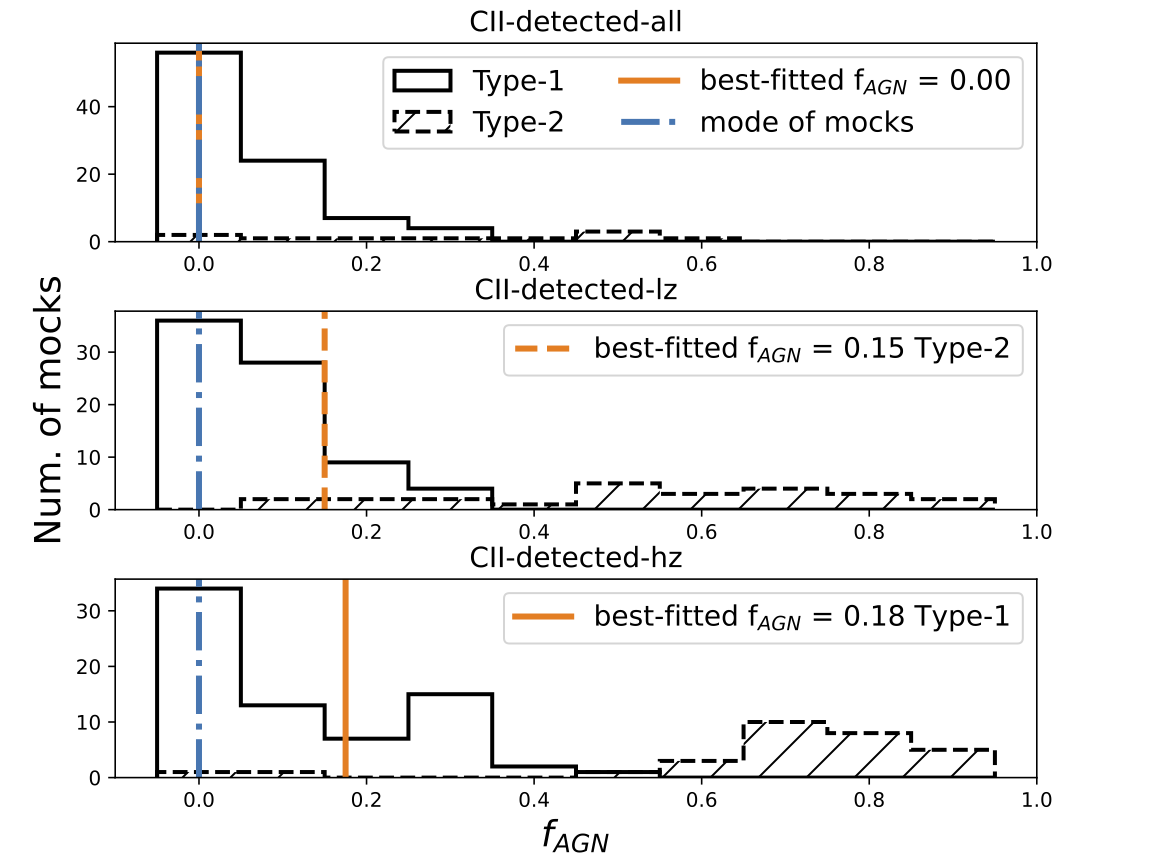}
    \caption{\textit{Top and bottom left: } The best-fitted SED model from \textsc{CIGALE} for the CII-detected-all, CII-detected-lz, and CII-detected-hz stacks from top to bottom, respectively. For each panel, the \textit{top} shows the observed photometric fluxes with errors (purple), the 3$\sigma$ upper limits fluxes (green triangles), the CIGALE-derived best model photometry (red dots), and the best-fitting CIGALE model (black). The best-fitting CIGALE model is the sum of contributions from an AGN (orange), dust-attenuated stellar emission (yellow; the intrinsic stellar emission is indicated in blue), nebular emission (green), dust emission (red) and radio nonthermal emission (brown). The \textit{bottom} shows the fractional discrepancies between the model and photometry. The reduced $\chi^2$ of best-fitted model are indicated in the top labels. \textit{Bottom right: } Histogram of $f_{\mathrm{AGN}}$ of CII-detected-all (\textit{top}), CII-detected-lz (\textit{middle}) and CII-detected-hz (\textit{bottom}) derived from 100 mocks, separated into Type-1 (solid histograms) and Type-2 (dashed histograms with hatch) based on their fitted viewing angle. Their best-fitted $f_{\mathrm{AGN}}$ are shown as orange vertical solid/dashed lines for Type-1/-2, respectively. The best-fitted AGN viewing angles are 70$^\circ$ and 30$^\circ$ for CII-detected-lz and CII-detected-hz, corresponding to Type-2 and Type-1 AGN, respectively. The blue dashed lines mark the statistical mode of the mocks. }
    \label{fig:sed}
\end{figure*}

In the bottom right panel of Figure \ref{fig:sed}, we show the histograms of $f_{\mathrm{AGN}}$ mocks separated into Type-1 and Type-2 AGN based on their fitted viewing angle, as well as the best-fitted $f_{\mathrm{AGN}}$ for the three sub-samples. 
The stacked SEDs of all three sub-samples are best-fit to little to no AGN activity and their mocks almost exclusively populate the lower $f_{\mathrm{AGN}}$ ($f_{\mathrm{AGN}}$ $\leq$ 0.2) regions. 
The statistical mode of all $f_{\mathrm{AGN}}$ is zero for mocks of all three sub-samples, suggesting that these sub-samples on average have little to no AGN activity. 
In addition, the CII-detected-lz/hz sub-samples are best-fitted to slightly higher $f_{\mathrm{AGN}}$, but consistent with that of the CII-detected-all sample within 1$\sigma$. 
Although, we note that the mocks of CII-detected-lz/hz span the full $f_{\mathrm{AGN}}$ range, which indicates that larger $f_{\mathrm{AGN}}$ values are possible, though still unlikely.

We note a caveat that the estimated $f_{\mathrm{AGN}}$ might be limited and potentially be biased low, due to the lack of near/mid-infrared constraint and the wavelength range (i.e., total IR) chosen by \textsc{cigale} to estimate the AGN fraction, where the contribution from galaxy might dilute any AGN contribution. 
Nevertheless, we find that, with the current photometry data, \textsc{cigale} can effectively recover the full $f_{\mathrm{AGN}}$ range for Type-1 AGN and high $f_{\mathrm{AGN}}$ values for Type-2 AGN, while under-estimate that of low-to-moderate Type-2 AGN by a factor of 2 (see more discussion in Appendix \ref{app:cigale_test}). 
Furthermore, the best-fitted $f_{\mathrm{AGN}}$ do not change significantly if the X-ray-detected galaxies is excised from the CII-detected-all and -hz sub-samples.

\subsubsection{Test on X-ray} \label{sec:xray_stack}

Secondly, we stack the X-ray observations, which might provide a more direct signal if AGN exist, especially for Type-1 AGN where X-ray emission is not obscured by a dusty AGN torus \citep{Padovani2017}. 
We use the publicly-available CSTACK4\footnote{CSTACK was developed by Takamitsu Miyaji and is available at \url{http:
//lambic.astrosen.unam.mx/cstack/}} tool to stack Chandra soft band ([0.5–2] keV) and hard band ([2-8] keV) X-ray images of galaxies in each sub-samples \citep{Miyaji2008}. 
We exclude the X-ray-detected galaxy in the CII-detected-all and -hz sub-samples. in order to diagnose the X-ray luminosity of those non-detection by stacking.  
None of sub-samples show a $>3\sigma$ detection in either bands.
Our stacked X-ray of CII-detected-all are down to uncertainties of $f_X = 2.1\times 10^{-17}$ erg s$^{-1}$ cm$^{-2}$ in [0.5–2] keV assuming a power-law X-ray spectrum with a slope of 1.4 and a galactic $N_{H}$ value of $2.6 \times 10^{20}$cm$^{-2}$.  The corresponding 3$\sigma$ upper limit of rest-frame X-ray luminosity is $L_{2-10\mathrm{keV}} = 10^{43.1}$ erg s$^{-1}$. These values are comparable to a moderate Seyfert, where $L_{2-10\mathrm{keV}} = 10^{42}$ erg s$^{-1}$ and $10^{44}$ erg s$^{-1}$ are the typical threshold to separate between SFGs/AGNs and Seyfert/quasar, respectively \citep{Marchesi2016b}. Adopting the bolometric correction of X-ray luminosity for radio-quiet AGN from \citet{Runnoe2012}, we obtain the 3$\sigma$ upper limit of bolometric luminosity $\sim10^{45.6}$ erg s$^{-1}$. However, we note that this bolometric correction may not well-quantified at L$_{2-10 keV} < 10^{43.5}$ erg s$^{-1}$, since only few sources in this range were included when deriving this bolometric correction. 
Thus, we are not able to make conclusion based on X-ray data.

\subsubsection{Test on Co-added Spectra} \label{sec:coadd_spectra}

We attempt one other approach here to probe the possible of AGN activity in the sample by utilizing the available rest-frame UV spectra from of the ALPINE galaxies from DEIMOS and VIMOS. In a study of similar redshift galaxies drawn from the VUDS survey, \citet{Nakajima2018} proposed diagnostic diagrams to classify the ionizing radiation field (star formation or AGN) of distant galaxies using the combination of C \textsc{iii}], C \textsc{iv}, and He \textsc{ii} lines (also see \citealp{Feltre2016, LeFevre2019}). They tested with a large grid of photoionization models and showed that these diagnostic diagrams can separate AGN and SFGs for a sample of C \textsc{iii}]-emitting galaxies at $z=2\sim4$ detected in VUDS at high level of purity and completeness. 
Unfortunately, however, none of the spectra for the galaxies in the ALPINE sample cover the wavelength of the C\textsc{iii}] feature and only galaxies in our lower redshift bin (z$\sim$4.5) have spectra that cover the CIV and HeII lines. Therefore, we combine spectra of CII-detected-lz subsample and measure the equivalent width (EW) of C\textsc{iv} and He \textsc{ii}. 

\begin{figure*}
    \centering
    \includegraphics[width=0.9\textwidth]{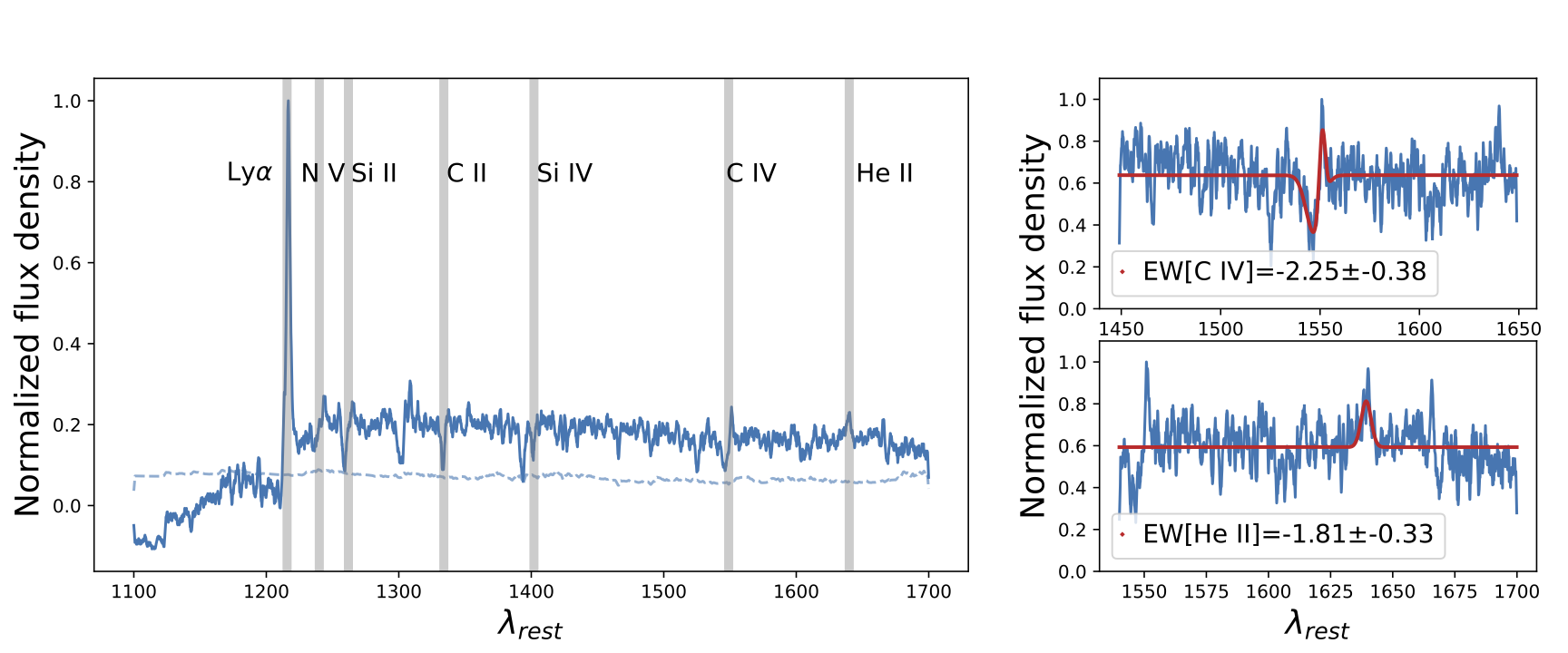}
    \caption{\textit{Left:} Inverse-variance, unit-weighted co-added DEIMOS spectra of 27 ALPINE galaxies in the CII-detected-lz sub-samples. The variances of the co-added spectra are shown as the dashed line. the Important spectral features are indicated by vertical dashed lines and labeled. \textit{Right:} zoom in of C \textsc{iv} and He \textsc{ii} lines. The best-fitted line profiles are shown as red lines with their EW values marked. The measured EW(C \textsc{iv}) versus EW(C \textsc{iv})/EW(He \textsc{ii}) fall in the SFGs region of the AGN/SFGs diagnostic diagrams of \citealp{Nakajima2018}.}
    \label{fig:spectra_coadd}
\end{figure*}

The spectra are combined (hereafter `co-added') through an inverse variance-weighted average after shifting each individual spectrum to the rest frame, interpolating on to a standard grid with constant plate scale of $\lambda = \lambda_{int}/(1+ z_{min})$, where $\lambda_{int}$ is the intrinsic plate scale specific to the instrument/setup, $z_{min}$ is the minimum $z_{spec}$ for each sample. Each spectrum is normalized to the average flux density in the rest-frame wavelength range $\lambda = 1350 -1400$. The intrinsic plate scales were set to 0.47$\AA$ for DEIMOS and 7.3$\AA$ for VIMOS. Due to these difference of the spectral resolution, DEIMOS and VUDS spectra are co-added separately. Note that we use the [C \textsc{ii}] as the systemic redshift $z_{spec}$ for each galaxy, as measurements with Ly$\alpha$ or ISM lines in the rest-frame UV do not necessarily probe the systemic redshift. 
For a total of 43 galaxies in the CII-detected-lz sub-sample, 42 galaxies have spectra with rest-frame wavelength coverage of $\lambda = 1200 - 1700$, including 27 DEIMOS and 15 VIMOS spectra, while one galaxy has DEIMOS spectra up to rest-frame $\lambda = 1450$, which is excluded in this exercise. 
We measure the EW of C \textsc{iv} and He \textsc{ii} following the method described in \citet{LeFevre2019}. Given that the p-cygni profile of C \textsc{iv}, we use a double Gaussian distribution to simultaneously fit the absorption and emission components. We measure the EW of He II using a single Gaussian model. 
The co-added spectra of DEIMOS is shown in Figure \ref{fig:spectra_coadd}, with zoom in of C \textsc{iv} and He \textsc{ii} shown on the left. The best-fitted line profiles are shown as red lines with their EW values marked in the bottom of panels. 
No strong C \textsc{iv} and He \textsc{ii} are observed. 
We derive an EW(C \textsc{iv}) = -2.23$\pm$-0.38  with the convention of negative EW indicating emission, and EW(C \textsc{iv})/EW(He \textsc{ii}) of 1.27 $\pm$ 0.27. These values and their uncertainties fall in the SFGs region of the diagnostic diagrams (see Figure 11 in \citealp{Nakajima2018}). 
In addition, no other AGN features are shown in the co-added spectra, such as the N \textsc{v} 1240 emission. 
For the VUDS coadded spectra, the CIV and HeII are not detected, and, thus we can draw no conclusions from these spectra other than these features are very weak, which indicates little to no AGN activity. 

To test whether one galaxy or small set of galaxies might be dominating the result despite the use of a normalization for all input spectra, we adopted a bootstrap analysis by randomly sampling the same number of galaxies as in our sample, with replacement, for the subset of galaxies with DEIMOS spectra and measure the EW of C \textsc{iv} and He \textsc{ii}. The bootstrap results are consistent with our measurements that these galaxies are not in the AGN region.

Overall, we do not find evidence of dominate AGN activity in any of our tests. 
Nevertheless, we note a caveat that we are not able to quantify the AGN contamination in the radio band, since AGN could contribute in radio band range over a factor of $10^5$ among different AGN types (e.g., \citealp{Panessa2019}). 
However, the main contribution to the radio emission in radio-quiet AGNs, at least up to z $\sim$ 1.5–2, is associated with star formation activity in the host rather than with radio jets \citep{Bonzini2015}.

\section{Discussion} \label{sec:discussion}

We recover lower $q_{\mathrm{TIR}}$ values of our sample than the local $q_{\mathrm{TIR}}$ relation for normal SFGs at $\sim3\sigma$ significance level. However, our $q_{\mathrm{TIR}}$ values are consistent with a variety of $q_{\mathrm{TIR}}$ for high redshift galaxies including pure normal SFGs, SFGs plus AGN, pure AGN, and SMGs.  
In section \ref{sec:fAGN}, we did not find evidence of dominant AGN activity with current multi-wavelength photometry in any of our sub-samples, and thus we dismiss the possibility of an AGN contribution as the dominant factor in lowering the $q_\mathrm{{TIR}}$. 
These results suggest that lower $q_\mathrm{{TIR}}$ values might be an intrinsic property of high-redshift (z $>$ 4) SFGs. 
In this section, we discuss the possible genesis for recovering $q_\mathrm{{TIR}}$ values that are lower than measured in local counterparts, in terms of the effect of different assumptions on the radio spectral index (Section \ref{sec:alpha}), the IR SED templates (Section \ref{sec:IRtemplates}) and astrophysical causes (Section \ref{sec:astrophysical}).

\subsection{The effect of the radio spectral index} \label{sec:alpha}

The radio emission of SFGs consists of thermal and non-thermal emission (see more in Section \ref{sec:intro}).  
The thermal radio emission has a typical power law spectrum  with $\alpha = -0.1$, and the non-thermal emission has a much steeper spectrum with $\alpha = -0.8$. The relative contributions of the two emissions vary with frequency. The non-thermal emission dominates the lower frequency ($<$5 GHz) and thermal emission may dominate at frequencies $>$10 GHz (see also \citealp{Price1992}). 
The fraction of thermal emission is found to be f$_{\mathrm{th, 1 GHz}} \sim $ 10\% at 1.4 GHz \citep{Condon1992, Niklas1997, Tabatabaei2017} and f$_{\mathrm{th, 10 GHz}} \sim$ 40\% at 10 GHz \citep{Gioia1982, Klein1988, Klein2018}. 
Our observation at 3 GHz corresponds to rest-frame 18 GHz at $z=5$. Thus, the observed 3 GHz flux might be dominated by thermal emission, rather than non-thermal emission dominated at rest-frame 1.4 GHz. 
If we assume a fraction of thermal emission of 10\% at 1 GHz and typical $\alpha = -0.1$ for thermal and $\alpha = -0.8$ for non-thermal spectrum, the spectra index between the rest-frame 18 GHz and 1.4 GHz is -0.60. 

As mentioned in Section \ref{sec:radio_detection}, the median $\alpha$ is found to be -0.8 for galaxies at $z > 2$ \citep{Delhaize2017}. Using $\alpha = -0.6$ and $\alpha = -0.8$, we re-calculate $q_{\mathrm{TIR}}$s for the three sub-samples, shown as blue open and shaded markers in the right panels of Figure \ref{fig:qTIR}. 
Employing these spectral indeces would slightly increase/decrease the $q_{\mathrm{TIR}}$ by 0.1 dex. Nevertheless, this relatively small change is not enough to explain our lower $q_\mathrm{{TIR}}$.

\subsection{The effect of the IR SED templates} \label{sec:IRtemplates}

Since only a single monochromatic FIR data point is available from our observations, our estimated IR luminosity depends largely on the choice of the IR SED template. Thus, we test whether a different IR SED template would change the IR luminosity enough to dramatically alter the recovered $q_{\mathrm{TIR}}$. 
As mentioned in \citet{Bethermin2020}, several works have constructed IR SED empirical templates up to $z \sim 4$ \citep{Alvarez2016, Bethermin2015, Bethermin2017, Schreiber2018, Alvarez2019}. 
Here we consider several other SED templates that might be representative of the average IR SED of our ALPINE galaxies and calculate the difference between the fiducial $L_{\mathrm{IR}}$ and the $L_{\mathrm{IR}}$ using our primary IR SED template. As a reminder, the primary IR SED template is adopted from \citet{Bethermin2015} with $\langle U \rangle = 50$ corresponding to the dust temperature T$_{\mathrm d} \sim$ 41K (see Section \ref{sec:sed} and Section \ref{sec:qTIR}). 
This is consistent with the T$_{\mathrm d}$ measured for four normal SFGs at $z=5.5$ using observations in three ALMA bands \citep{Faisst2020b}. 
The changes of $q_{\mathrm{TIR}}$ when using different templates are summarized in Table \ref{tab:IRSED}.

First, we consider other possible SED templates from \citet{Bethermin2015}. In detail, \citet{Bethermin2015} measured the evolution of the average SED by varying the $\langle U \rangle$ parameter and using a sample of massive ($>3\times10^{10}$) main-sequence SFGs and starbursts in the COSMOS field. They found that the stacked SED of main-sequence SFGs in the redshift range of $3.5 < z < 4.0$ is best fitted to the SED template with $\langle U \rangle = 72.98$. 
Such a template yields a slightly larger fiducial $L_{\mathrm{IR}}$, increasing the primary $L_{\mathrm{IR}}$ by 0.07 dex.  

Second, we consider six possible SED templates constructed by \citet{Schreiber2018}. They presented a SED library characterized by the dust mass and the mid-to-total infrared color (IR8). Applying this library to SFGs at $0.5 < z < 4$ in the deep CANDELS fields, using both individual detections and stacks of Herschel and ALMA imaging, they found trends of increasing $\mathrm{T_d}$ and IR8 with redshift and distance from the main sequence. 
We include the best-fitted SED template of SFGs at their highest redshift bin ($3.5 < z < 5.0$) that is characterized by $\mathrm{T_d}$ of 41.8 K and IR8 of 7.37. 
Following the $\mathrm{T_d}$ and IR8 relation (eq. 15 and 16 in \citealp{Schreiber2018}), the $\mathrm{T_d}$ are found to be 44.4 K and 49.0 K at z = 4.5 and z = 5.5, and a fixed IR8 = 7.37. 
In addition, \citet{Schreiber2018} found in general higher IR8 values for low mass galaxies ($M_* < 10^{10} M_\odot$) up to $z \sim 2$, which is comparable with the median stellar mass of the CII-detected-all sub-sample ($10^{9.78}$ M$_\odot$). 
Thus, we also include templates with a higher IR8 (IR8 = 10.7) that are best-fitted to low mass galaxies at $1.2 < z < 1.8$.  
Using the combination of three $\mathrm{T_d}$ and two IR8 values, we re-calculate the $q_{\mathrm{TIR}}$ and obtain an average increase of 0.2 dex in IR luminosity (see the change using each template in Table \ref{tab:IRSED}).  
Thus, we add an additional 0.2 dex in our $L_{\mathrm{IR}}$ values and plot the corresponding fiducial $q_{\mathrm{TIR}}$ as orange open markers in the right panel of Figure \ref{fig:qTIR}.

Thirdly, we consider the SED library constructed by \citet{Alvarez2019} using the CIGALE SED fitting code and fitted to a large sample of Lyman break galaxies at redshift $2.5 < z < 3.5$ in the COSMOS field, binned in terms of stellar mass, UV luminosity ($L_{FUV}$), and UV continuum slope ($\beta$). 
We use the best-fitted SED template for their lower mass galaxies (LBG-M1: log(M$_*$/M$_\odot$) = 9.75 - 10.00 and LBG-M2: log(M$_*$/M$_\odot$) 10.00 - 10.25).
We calculate a smaller fiducial $L_{\mathrm{IR}}$, decreasing the primary $L_{\mathrm{IR}}$ by 0.59 dex and 0.52 dex. This might be due to these templates having lower $\mathrm{U_{min}}$ ($\mathrm{U_{min}}$ = 30.9 and 36.9 for the two lower mass bins), which corresponds to a lower $\mathrm{T_d}$ as compared to templates in \citealp{Schreiber2018} and \citet{Bethermin2015}. To show the change of $q_{\mathrm{TIR}}$ when using these IR SED templates, we reduce 0.5 dex in our $L_{\mathrm{IR}}$ values and plot the corresponding fiducial $q_{\mathrm{TIR}}$ as orange shaded markers in the right panel of Figure \ref{fig:qTIR}.

Finally, we consider the SED template constructed by \citet{DeRossi2018} based on the SED of Haro 11, a local moderately low metallicity galaxy undergoing a very young and vigorous starburst that could be similar to the conditions of high redshift galaxies. They showed that high level of consistency with the Haro 11 SED template and the measurements of individual galaxies at $z > 5$ with adequate FIR observations. We obtain a smaller fiducial $L_{\mathrm{IR}}$ decreasing the primary $L_{\mathrm{IR}}$ by 0.4 dex.

We find that the estimated L$_{\mathrm{IR}}$ depends on the adopted IR SED template and the assumption of dust temperature. Applying different IR templates can change the $q_{\mathrm{TIR}}$ more significantly and place the tension between our stacked $q_{\mathrm{TIR}}$s and those of local SFGs at the $<1\sigma$ significance level.

\begin{deluxetable*}{lll}
\tablecaption{Possible IR SED templates and the change on the recovered $q_{\mathrm{TIR}}$ \label{tab:IRSED}}
\tablewidth{0pt}
\tablehead{
\colhead{Templates} & \colhead{Criteria} & $\Delta q_{\mathrm{TIR}}$ [dex]\\
(1)& (2) & (3) }
\startdata
 \citet{Bethermin2015} & $\langle U \rangle = 50$ & \\
 \citet{Bethermin2015} & $\langle U \rangle = 72.98$ & + 0.07\\
 \citet{Schreiber2018} & $\mathrm{T_d} =$ 41.8 K, 44.4 K, 49.0 K and a fixed IR8 = 7.37 & +0.12, +0.19 and +0.34\\
 \citet{Schreiber2018} & $\mathrm{T_d} =$ 41.8 K, 44.4 K, 49.0 K and a fixed IR8 = 10.7 & +0.07, +0.15 and +0.30\\
 \citet{DeRossi2018}   & Haro 11 & -0.4 \\
 \citet{Alvarez2019}   & $\mathrm{U_{min}}$ = 30.9 and 36.9 & -0.59 and -0.52 \\
\enddata
\end{deluxetable*}

\subsection{Astrophysical causes} \label{sec:astrophysical}

As mentioned previously, the radio emission in the SFGs directly originates from the star formation process and is expected to reflect the star formation rate on a relatively short time scale. However, the IR emission is due to the energy from star formation that is absorbed and re-emitted by dust, which depend on the amount of dust, quantified as the dust attenuation. 
Thus, if IR and radio emission purely originate from star formation, the IR-Radio correlation should be proportional to the dust attenuation. 
If the dust attenuation of ALPINE galaxies is lower than that of local SFGs, it would depress the IR emission and result in a lower $q_{\mathrm{TIR}}$ when compared to that of local SFGs. 

Indeed, \citet{Fudamoto2020} have studied the dust attenuation of ALPINE galaxies, in terms of the ultraviolet (UV) spectral slope ($\beta$), M$_*$, and infrared excess (IRX = $L_{\mathrm{IR}}/L_{\mathrm {UV}}$). They found lower IRX values for individual ALPINE galaxies and also stacks, implying a lower obscured fraction of star formation than at lower redshift. 
\citet{Fudamoto2020} measured the obscured fraction of star formation ($f_{\mathrm{obs}} = \mathrm{SFR_{IR} / SFR_{total}}$) by stacking ALPINE galaxies in different stellar mass and redshift bins. For $M_* > 10^{10} M_\odot$, the stacked $f_{\mathrm{obs}}$ is $0.67^{+0.05}_{-0.07}$ at $z\sim4.5$ and $0.44 \pm 0.11$ at $z\sim5.5$, as compared to $f_{obs} > 0.80$ at $z = 0 - 2.5$ at the same stellar mass. For lower mass $M_* < 10^{10} M_\odot$, the $f_{\mathrm{obs}}$ is much lower with $0.36^{+0.13}_{-0.19}$ at $z\sim4.5$ and $3\sigma$ upper limit of $f_{\mathrm{obs}} < 0.43 $ at $z\sim5.5$, similar to $f_{\mathrm{obs}}$ at z = 2.5 but lower than $\sim$0.6 at z=0. 
It is possible that the dust buildup, which primarily governs the IR emission in addition to older stellar populations, does not have enough time to occur fully in these galaxies, whereas the radio emission can happen on a more rapid timescale. 
For simplicity, we assume a $f_{\mathrm{obs}} = 0.5$ for our ALPINE galaxies and $f_{\mathrm{obs}} = 0.8$ at z = 0. 
We find that, under these assumptions, the $q_{\mathrm{TIR}}$ values increase by 0.2 dex, as shown as green open markers in the right panel of Figure \ref{fig:qTIR}, which places the tension between our stacked $q_{\mathrm{TIR}}$s and those of local SFGs at the $<1\sigma$ significance level. 

Nevertheless, slightly brighter normal SFGs at high redshift might have similar obscured fraction as local SFGs. \citet{Bowler2021} measured the dust continuum of five normal SFGs at $z\sim7$, and found that the IRX-$\beta$ in these galaxies are consistent with the IRX-$\beta$ relation of local star-burst galaxies. However, these galaxies are three times brighter in rest-frame UV than our ALPINE galaxies on average. Thus, their obscured fraction might be different from our sample.

There are some other possible explanations for the lower $q_{\mathrm{TIR}}$. 
Recently, \citet{Algera2020} have shown that either increasing the magnetic field or decreasing the density of ISM by a factor of five could decrease the $q_{\mathrm{TIR}}$ by 0.2 dex up to $z \sim 4$. 
In addition, \citet{Bressan2002} found that $q_{\mathrm{TIR}}$ is dependent on the stage of evolution of the stellar populations in the galaxy.
They suggested that, in the post-starburst phase, non-thermal synchrotron emission dominates, thus increasing the apparent radio flux. 
By modeling different star formation histories with e-folding time scales in the range of 10 - 50 Myr, they found that, in all models, $q_{\mathrm{TIR}}$s decrease with time and reach $q_{\mathrm{TIR}} <2 $ after $\sim$25 Myr and remain at $q_{\mathrm{TIR}} <2$ for another $\sim$20 Myr. 
Although, we are not able to quantitatively test these hypothesis with our data, due to the lack of information such as ISM density, magnetic field and stellar age, it is plausible that the ISM densities are higher in ALPINE galaxies, since galaxies are found to be more compact at high-$z$ \citep{vanDokkum2010, VanderWel2014} and contain larger ISM gas content at high-$z$ \citep{Santini2014}.

\section{Summary} \label{sec:summary}

We have studied the radio properties of 66 galaxies in the COSMOS field that have [C \textsc{ii}] detections from the ALPINE survey by exploiting the stacking technique. 
We separated them into three sub-samples depending on their spectroscopic redshift named CII-detected-all, -lz and -hz. 
We detected a radio signal of CII-detected-all and -lz in their median-stacked 3 GHz image and place meaningful limits on the CII-detected-hz sub-sample. 
Our main conclusions are as follows:

\begin{itemize}

    \item We recovered lower $q_{\mathrm{TIR}}$ values of our stacked samples at $z\sim4.6$ than the local $q_{\mathrm{TIR}}$ relation for normal SFGs at $\sim3\sigma$ significance level. However, our $q_{\mathrm{TIR}}$ values are broadly consistent with that of SMGs at $2 < z < 5$ and all galaxies including SFGs plus AGN and pure AGN at z$\sim$3. 
    
    \item While powerful broad-line AGN were selected against when constructing the ALPINE sample, the prevalence of lower luminosity and/or obscured AGN was previously unknown. Although, the data are not sufficient to confirm whether AGN contribute to the radio band either through spectral index or SFR estimates, our samples do not exhibit evidence of dominant AGN activity in the stacked optical-to-far-IR SED, X-ray and UV spectra, which rule out the possible of the AGN component as the main contributor to the observed deviation.  The stacked SEDs of all three sub-samples are best-fit to little to no AGN activity and their $f_{\mathrm{AGN}}$ mocks dominate the lower $f_{\mathrm{AGN}}$ ($f_{\mathrm{AGN}} \leq 0.2$) regions. The co-added UV spectra of CII-detected-lz show no evidence of AGN activity. The upper limits of $L_{\mathrm{AGN}}$ from stacked SEDs and X-ray are $< 10^{45.6}$ erg s$^{-1}$. 
        
    \item We explore various effects that might reduce the tension between our stacked $q_{\mathrm{TIR}}$ values and those of local SFGs, including changing fiducial radio spectral index and applying different IR templates. Either of them can reduce the tension to the $<1\sigma$ level.

    \item In addition, such tension can be alleviated based on the fact that lower obscured fraction of star formation has been found in ALPINE galaxies than that in local galaxies and even galaxies at $z\sim2$ \citep{Fudamoto2020}. It is possible that the dust buildup, which primarily governs the IR emission in addition to older stellar populations, has not had enough time to occur fully in these galaxies, whereas the radio emission can respond on a more rapid timescale. Thus, we might expect the IR-radio correction to be modified at high redshift. 

\end{itemize}

Future observations of a large sample of normal high redshift SFGs are essential to test whether the lower $q_{\mathrm{TIR}}$ are ubiquitous in these galaxies. Indeed, we expect that future JWST projects with observation at mid-infrared, such as CEERS and PRIMER, could provide stronger constraint on the dust obscured star formation and AGN activities for high-redshift galaxies. 
In addition, we expect that combined deeper radio and FIR observations in more than one band will provide a robust way to calculate the $q_{\mathrm{TIR}}$ and to study the evolution of IRRC.

\section*{Acknowledgements}

LS acknowledges the National Natural Science Foundation of China (NSFC) No. 12003030 and the Fundamental Research Funds for the Central Universities. 
GL acknowledges the research grants from the China Manned Space Project (No. CMS-CSST-2021-A06 and No. CMS-CSST-2021-A07), the NSFC No. 11421303. 
WF acknowledges the NSFC Grants No. 11773024. 
This work is also based in part on observations taken by the 3D-HST Treasury Program (GO 12177 and 12328) with the NASA/ESA HST, which is operated by the Association of Universities for Research in Astronomy, Inc., under NASA contract NAS5-26555. 
Part of the material presented herein is based upon work supported by the National Aeronautics and Space Administration under NASA grant no. 80NSSC21K0986.
This material is based upon work supported by the National Science Foundation under Grant No. 1908422. 
M.B. gratefully acknowledges support by the ANID BASAL project FB210003 and the FONDECYT regular grant 1211000.
G.C.J. acknowledges ERC Advanced Grants 695671 ``QUENCH’’ and 789056 ``FirstGalaxies’’, as well as support by the Science and Technology Facilities Council (STFC). 
M.R. acknowledges support from the Narodowe Centrum Nauki (UMO-2020/38/E/ST9/00077).
E.S. thanks the LSSTC Data Science Fellowship Program, which is funded by LSSTC, NSF Cybertraining Grant \#1829740, the Brinson Foundation, and the Moore Foundation.
E.I. acknowledge funding by ANID FONDECYT Regular 1221846. 

This paper is based on data obtained with the ALMA Observatory, under Large Program 2017.1.00428.L. ALMA is a partnership of ESO (representing its member states), NSF(USA) and NINS (Japan), together with NRC (Canada), MOST and ASIAA (Taiwan), and KASI (Republic of Korea), in cooperation with the Republic of Chile. The Joint ALMA Observatory is operated by ESO, AUI/NRAO and NAOJ. 
This study is based on data taken with the Karl G. Jansky Very Large Array which is operated by the National Radio Astronomy Observatory. The National Radio Astronomy Observatory is a facility of the National Science Foundation operated under cooperative agreement by Associated Universities, Inc. 
This work is based in part on observations made with the Spitzer Space Telescope, which is operated by the Jet Propulsion Laboratory, California Institute of Technology under a contract with NASA. 
Furthermore, this work is based on data from the W.M. Keck Observatory, the Canada-France-Hawaii Telescope, the Subaru Telescope and retrieved from the HSC data archive system, which is operated by the Subaru Telescope and Astronomy Data Center at the National Astronomical Observatory of Japan. 
We wish to thank the indigenous Hawaiian community for allowing us to be guests on their sacred mountain, a privilege, without with, this work would not have been possible. We are most fortunate to be able to conduct observations from this site.


\appendix 
\section{Optical/IR/Radio Stacks} \label{app:allstackes}

\begin{figure*}
    \centering
    \includegraphics[width=0.9\textwidth]{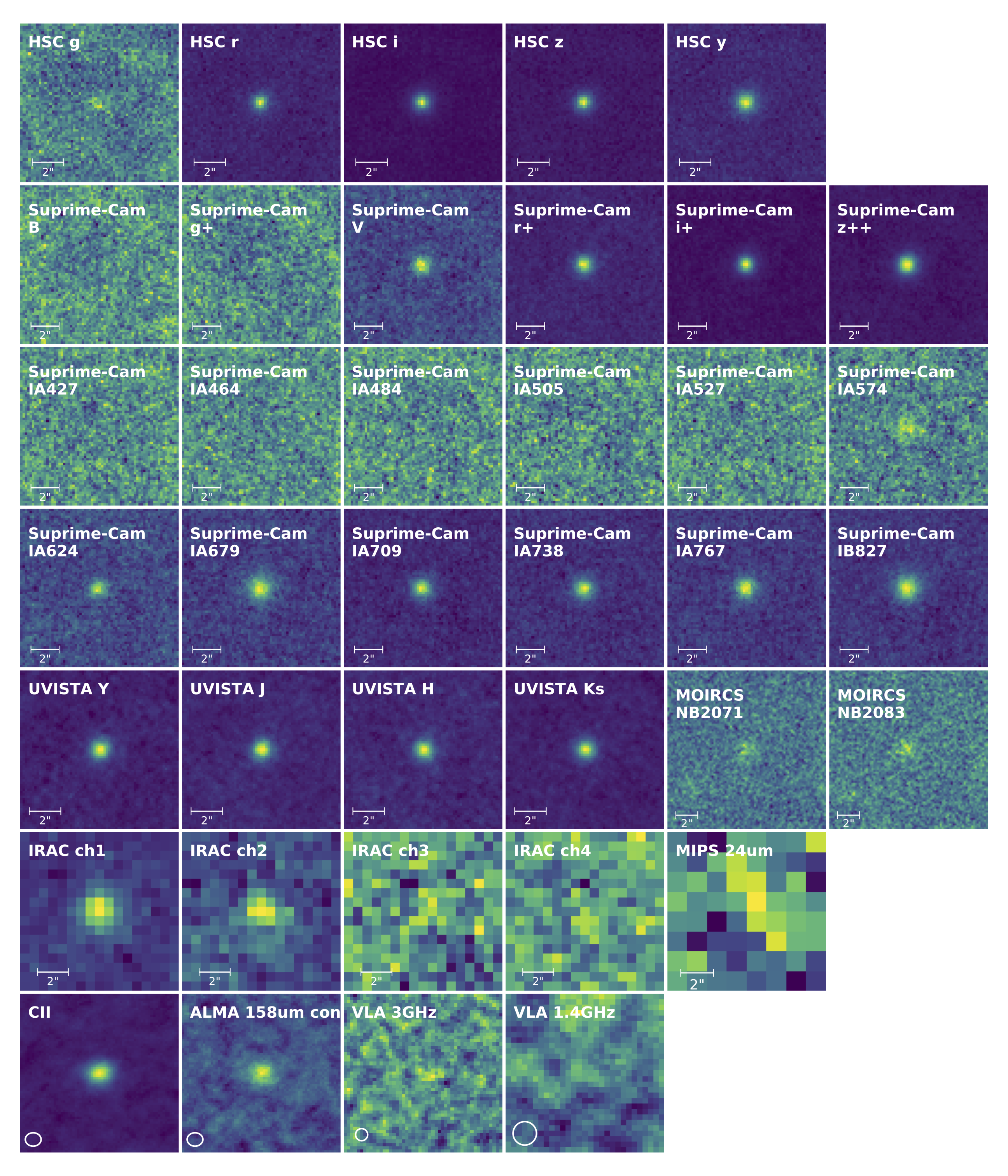}
    \caption{The $10\arcsec \times 10\arcsec$ stack cutouts of all available bands for the CII-detected-all sample. The wavelength generally increases from top left to bottom right. The instrument and band are listed for each cutout. See Section \ref{sec:stacking} for the stacking method.  }
    \label{fig:allstacks}
\end{figure*}

To illustrate the data quality of stacks, in Figure \ref{fig:allstacks}, we show the stack cutouts of all available bands for the CII-detected-all sample. 

\section{Test on \textsc{cigale} SED fitting with current photometry set} \label{app:cigale_test}

We use the CIGALE SED fitting to rule out the possibility of considerable obscured and unobscured AGN activity among the ALPINE sample. We applied mock analysis in CIGALE in order to check the reliability of estimated parameters. However, due to the upper limits given at 5.8$\mu$m, 8$\mu$m and 24$\mu$m, the SED constraint on AGN activity is limited, especially for the Type-2 AGN, which is also indicated by the large uncertainties on $f_\mathrm{AGN}$. 

To further test the efficacy of CIGALE SED fitting in constraining AGN activity with the current photometry set,  a grid of synthetic models are created by CIGALE using the best-fitted parameters for the stacked SED of the CII-detected-all sample. For this test, the $f_\mathrm{AGN}$ is allowed to vary between 0 - 0.9 and the viewing angle is set to $\theta$ = 30$^{\circ}$ for Type-1 and $\theta$=70$^{\circ}$ for Type-2. For each synthetic model, we generate 100 mock sources by assigning mock fluxes based on the fluxes from the model and using a signal-to-noise ratio (SNR) equivalent to that of the corresponding stacked photometry. In detail, mock flux densities in a given band are assigned by initially randomly sampling a Gaussian distribution with mean equal to the model flux and error appropriate for the SNR of a given data point in the stacked photometry, effectively reproducing a photometric measurement of the model spectrum. 
For those bands where only upper limits are available (e.g., IRAC channel 3, 4, and MIPS 24$\mu$m), we set the mock flux densities in these bands by multiplying the model flux density by 40-90, in order to match the ratio of $3\sigma$ upper limits and best-fitted model values ($\sim$86, 87, 47 for IRAC channel 3, 4 and MIPS 24$\mu$m, respectively). The $f_\mathrm{AGN}$ of Type-1 AGN are well recovered in the full $f_\mathrm{AGN}$ range, with small uncertainty ($\sim$ 0.1). 
While, those of Type-2 have larger uncertainties and are slightly offset to lower fAGN values for fAGN $<=$ 0.5. These results suggest that, with the current photometry set, CIGALE can robustly recover $f_\mathrm{AGN}$ for Type-1 AGN in the full $f_\mathrm{AGN}$ range and Type-2 AGN in the high $f_\mathrm{AGN}$ range, while, for Type-2 sources with low-to-moderate $f_\mathrm{AGN}$, $f_\mathrm{AGN}$ could be underestimated by a factor of two.



\end{document}